\documentclass[11pt,a4paper]{article}

\usepackage[
  left=1.5cm,
  right=1.5cm,
  top=2cm,
  bottom=2.5cm
]{geometry}
\usepackage{graphicx}
\usepackage{amsmath,amssymb}
\usepackage{authblk}
\usepackage{hyperref}
\usepackage{braket}

\usepackage{nccmath}

\usepackage{cases}
\usepackage{comment}
\usepackage{placeins}
\RequirePackage{pgf}
\usepackage{fixmath}
\newlength{\extralength} 
\usepackage{tikz}
\usepackage{colortbl}
\usetikzlibrary{arrows,decorations.pathmorphing,backgrounds,fit,positioning, decorations.pathreplacing, calc,shapes,angles,quotes, decorations.markings}
\usetikzlibrary{backgrounds}   

\usepackage{pgfplots}
\pgfplotsset{compat=1.18} 
\usetikzlibrary{calc}

\title{Understanding Squeezed States of Light Through Wigner's Phase-Space}

\author[1]{Sibel Ba{\c s}kal}
\author[2]{Marilyn E. Noz}
\affil[1]{Department of Physics, Middle East Technical University, 06800 Ankara, Türkiye}
\affil[2]{Department of Radiology, New York University, New
  York, NY 10016, USA}
\date{}

\begin{document}
\maketitle

\begin{abstract}
This paper starts with the transition from classical physics to quantum mechanics which was greatly aided by the concept of phase space.  The role of canonical transformations in quantum mechanics is addressed. The Wigner phase-space distribution function is then defined which arises from the formulation of the density matrix, followed by the harmonic oscillator in phase space. 
Coherent and one- and two-mode squeezed states of light as well as the squeezed vacuum are discussed in the phase-space picture. Attention is also drawn to the fact that squeezed states naturally generate entanglement between the two-modes. Coupled harmonic oscillators are also elucidated in connection with the Wigner phase space. It will be noted that the phase-space picture of quantum mechanics has become an important scientific language for the rapidly expanding field of quantum optics. Here, we mainly focus on the simplest form of the Wigner function, which finds application in many branches of quantum mechanics. We make use of several symmetry groups such as Lorentz groups, the symplectic group in two and four dimensions, and the Euclidean group. The decoherence problem of an optical field is examined through a reformulation of the Poincaré sphere as a further illustration of the density matrix.
\end{abstract}

Keywords: Wigner's phase-space; Wigner function; density matrix;  Wigner function of harmonic oscillator; canonical transformations; coherent states;  single-mode squeezed state; squeezed vacuum; two-mode squeezed states of light; decoherence and the Poincaré sphere

\section{Introduction}\label{sec:intro} 
In this paper we shall review in a unified framework the concepts of quantum phase space, canonical transformations, the density operator,
the Wigner distribution function and harmonic oscillator in phase space, in order to
elucidate squeezed states of light using the Wigner phase space distribution function. The approach we have taken, namely using the Wigner phase-space distribution function, should provide a better understanding of how these tools in quantum optics are coordinated to form a unified picture. In particular, the backbone of this unification is provided by a group theoretical treatment.

We start with the Hamiltonian formalism of classical mechanics that forms the origin of the
concept of phase space in which a dynamical system depends
on both independent coordinate variables and momentum variables. Phase space thus consists of $2n$ Cartesian coordinate variables~\cite{goldstein80}.
This phase-space formalism is the starting point for
the modern approach to classical mechanics~\cite{arnold97,abra78}, which includes nonlinear dynamics and chaos~\cite{xu_2017}. 
Phase space in quantum mechanics is explored in the books~\cite{knp91,zachos_2005}. The phase space formalism is quite different from that of the Schrödinger and Heisenberg formulations.

Linear canonical transformations correspond to unitary transformations
in Schrödinger quantum mechanics. The symplectic group~\cite{weyl46} provides the mathematics for linear canonical transformations.
Homogeneous linear canonical transformations in phase space can consist 
of $n$ pairs of canonical variables which 
are governed by the group
$Sp(2n)$~\cite{gilmore74,guil84}.
In this paper, physical problems
requiring one and two pairs of canonical variables will be our primary concern.
The symmetry group for one pair of canonical variables is the inhomogeneous symplectic group commonly
denoted by $ISp(2)$, which has the two-dimensional homogeneous symplectic group $Sp(2)$ as a subgroup. Included also is the two-dimensional Euclidean group $E(2)$, and the Lorentz group $SO(2\, ,1)$ which is locally isomorphic to the $Sp(2)$ group.
The four-dimensional phase space has two pairs of canonical variables and is governed by the (3 + 2)-dimensional de Sitter (Lorentz $SO(3\, ,2)$) group that is locally isomorphic to the $Sp(4)$ group. All of the groups mentioned above are Lie groups. 

\begin{figure}[htb]
\centering
\begin{tikzpicture}[scale=0.90, transform shape]



\draw[ultra thick, blue, fill=blue!20] (3,3) circle (1cm);

\draw[ultra thick,red, fill=red!20,rotate around={45.0:(3,3)}] (3,3) ellipse (3 cm and 0.333 cm);

\path (6.2,3) node [black, scale=1.1] {$x$};
\path (3,6.7) node [black, scale=1.1] {$p$};

\draw[black,  thick, ->, >=stealth] (0,3) -- (6,3);
\draw[black,  thick,->, >=stealth] (3, 0.5) -- (3, 6.5); 

\draw[thick, <->, stealth-stealth, color=blue] (2, 1) -- node (M4)[fill=white,midway, scale=0.6] {\huge$\Delta x$} (4.0, 1);
\draw[thick, <->, stealth-stealth, color=blue] (5.0, 2) -- node (M5)[fill=white,midway, scale=0.6] {\huge$\Delta p$} (5.0, 4);
\draw[color=red] (5.2, 5.75) -- node (M6)[fill=white,midway, scale=0.6] {\huge$\Delta x$} (5.7, 5.25);
\draw[thick, |<-, >=stealth, color=red] (5.25, 5.75) --(4.75, 6.25) ;
\draw[thick, |<-, >=stealth, color=red] (5.75, 5.25) -- (6.25, 4.75);
\draw[thick, |<->|, >=stealth, color=red] (-0.5, 1.75) -- node (M7)[fill=white,midway, scale=0.6] {\huge$\Delta p$} (4.0, 6.1);
\end{tikzpicture}
\caption{Illustrated as a blue circle, it corresponds to the initial vacuum state in phase space, 
whose area is proportional to and is a representative of minimum uncertainty $\Delta~x\Delta~p= 1/4$.
When the circle is squeezed,  it is transformed into  an ellipse, shown in red. This has the same area as the circle and thus 
is also a representative of minimum uncertainty.} \label{fig:11}
\end{figure}
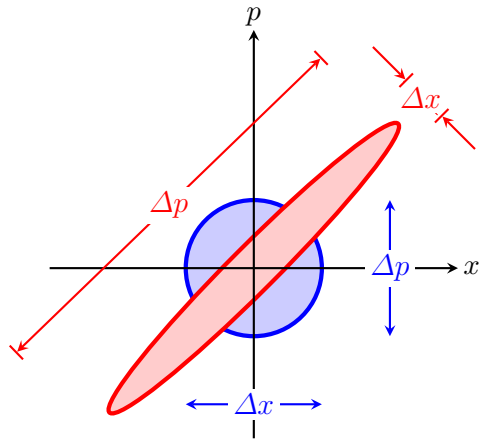

Similar to the wave function playing the central role in the
Schrödinger picture, the starting point in the
phase-space picture of quantum mechanics is provided by the  distribution function introduced by Wigner~\cite{wig32a}. 
Linear canonical transformations in phase space are directly applicable to the Wigner function and correspond to unitary transformations in the Schrödinger picture of quantum mechanics.
Strictly defined within the framework of quantum mechanics, the Wigner distribution function is defined in phase space where both $x$ and $p$ are c-numbers. Since in quantum mechanics, the position and momentum variables cannot be simultaneously measured, the problem of how to represent the uncertainty principle in this picture must be answered. 
In optics, and particularly in quantum optics, the associated physics can be described by using the Wigner phase space. We let the $x$-axis and the $p$-axis be orthogonal to each other. In this context, Heisenberg’s uncertainty relation becomes $\Delta~x\Delta~p \geq 1/4$ with $\Delta~x$ and $\Delta~p$ representing uncertainties in the two quadratures. Then in terms of uncertainties we present the area of a circle centered around some $(x,p)$. When we "squeeze" the circle it becomes an ellipse while preserving the area. This is shown pictorially in Figure~\ref{fig:11}. The fact that uncertainty in one quadrature can be reduced at the expense of increased uncertainty in the other quadrature is one of the most prominent features of squeezed light. We shall discuss this further in terms of the squeezed states of light in Section~\ref{sec:9}.

In formulating phase space, the Wigner phase-space distribution function is very closely related to the density matrix.
The density matrix formalism is a
convenient representation for pure and mixed (non-pure) states and is especially important when not all measurable variables are measured in laboratories. Since the time when von Neumann and separately, Landau introduced the statistical (density) operator in~\cite{Neumann_1927,neumann1927b,landau_1927}.  Dirac's contribution ensured that the concept is well understood~\cite{dirac_1930}. Since then there have been many books~\cite{blum_2012} and articles~\cite{fano_1954} dedicated to the subject of the density matrix.
Because the density matrix and the Wigner function are directly related, the Wigner function also addresses various treatments of the density matrix. Photonics and classical-wave analogs, coherency (polarization) matrices, and Stokes parameters for partially coherent light rely on the density matrix, as does radio-astronomy interferometry and adaptive optics mirrors. We shall discuss some of this in Section~\ref{sec:14.3}.

Phase space is extensively employed to advance knowledge in the realms of quantum optics~\cite{leonhardt_1997, schleich_2001,Simon_2000,colas_2022}, 
quantum computing~\cite{aad_search_2023,mari_2012,bianucci_discrete_2002}, quantum communication~\cite{miquel_quantum_2002},
in a wide range of two-mode quantum models as quantum information~\cite{sanchezsoto_2025}, 
in condensed matter~\cite{ng_phase-space_2019}, many-body physics~\cite{liang_floquet_2018},
high energy physics~\cite{yuan_2010,mahlein_2023}, nuclear physics~\cite{baker_1960}, collisions in atomic physics~\cite{lee_1995,kurtsiefer_1997}, and decoherence~\cite{brody_phase-space_2025,oconnell_2003,chun_2003,zurek_2003,zurek_sub-planck_2001}. In the context of quantum information the focus is on discrete quantum systems, where an associated Wigner function is introduced pertinent to the system described~\cite{rundle_2021}. The symmetry properties of those systems rely on the groups $SU(2)$~\cite{sanchezsoto_2025} and $SU(1\,,1)$~\cite{seyfarth_2020}.

We note that the Wigner function that we particularly focus on in this article is not the only phase
space distribution function.
There are equivalent ways to represent the same density operator $\hat\rho$
differing only by operator ordering.  For instance when considering the
Cahill–Glauber $s-$ordered quasiprobabilities $W(s)(\alpha), s\, \epsilon\, [-1,1]$ (where $s$
encodes the ordering choice), we have these three well-known examples~\cite{cahill_1969,gerry_2024}: \\
$i. \,s=+1: \,P(\alpha)\, \mbox{(normal ordering)}, \\
ii. \, s=0:  W(\alpha) \, \mbox{Wigner (Weyl / symmetric ordering)}\\
iii.\, s=-1: Q(\alpha) \, \mbox{(antinormal ordering)}$\\
where $P(\alpha)$ stands for Glauber–Sudarshan and $Q(\alpha)$ stands for Husimi
phase-space distributions.
We shall not go into the details of these probability distribution
functions, however an interested reader can see the following recent references
in that regard~\cite{linowski_2024,andreev_2011}. 

The quantum information field arising from quantum optics has extremely profited from utilizing continuous variable (CV) quantum information carriers as an alternative for qubits~\cite{weedbrook2012}. It is observed that CV photonics is highly effective in generating many entangled modes by means of producing and using many distinct squeezed modes of light. The most familiar example of continuous quantum information is the quantized harmonic oscillator, which can be described by continuous variables, namely the position and
momentum variables~\cite{braunstein_2005, braunstein_2003}, as we shall be discussing in this article. There are many theoretical papers that study this topic by using Wigner functions~\cite{weedbrook2012} as well as those from an experimental standpoint 
~\cite{rahman_local_2025,fluhmann_encoding_2019}. 
Experimental approaches extend to quantum error correction and deterministically encode quantum information which employ superconductivity as well as harmonic oscillators~\cite{campagne-ibarcq_quantum_2020,vlastakis_2013} and to another that employs superconductivity and Wigner tomography along with superposition to measure arbitrary quantum states~\cite{hofheinz_synthesizing_2009}.

The theoretical framework of quantum optics is based on coherent and squeezed states of light and begins with the creation and annihilation operators, which consist of the
harmonic oscillator step-up and step-down operators, respectively, in Fock space. Unitary transformations produced by the linear form of these operators
in the vacuum state generate coherent states.  
When the transformation is generated by the quadratic form of these creation and annihilation operators a squeezed state results, and when the
quadratic form involves two different photons a two-mode squeezed state is generated. In the Schrödinger picture of quantum mechanics the theory of coherent and squeezed states can be considered as essentially an algebra of creation and annihilation operators applied to harmonic-oscillator states in Fock space.
In that space generalized multi-mode squeezed states can also be 
constructed~\cite{Lo_1993,simon_1994,ma_1995}. It is also claimed that higher mode squeezed states can be viewed as an effective approach to improve entanglement~\cite{dai_2025}. By using canonical transformations in quantum phase space every coherent or squeezed state can be generated from the vacuum state. Consequently, their symmetry properties are retained as they are inherited from the vacuum through these transformations.

We make a few definitions before we start that will be useful as we continue with the main body of the paper. 
Unless explicitly stated, we use $c=\hbar=1$ . We also note that the trace and the determinant operation result in c-numbers.
A Lie group~\cite{gilmore74}, is a group that is also a smooth manifold, 
where the group operations, specifically multiplication and inversion, are smooth maps. 
The properties of Lie groups are determined by a closed set of generators, which are known as the Lie algebra of the group. 
The Lie algebra can be obtained from the Lie group by exponentiation. 
If $X$ is considered as any complex-valued square matrix, then
\begin{equation}\label{eq:023}
    \exp X = \sum ^{\infty}_{n=0} \frac{1}{n!} X^{n} \, .
\end{equation}
The group element that corresponds to any parameter $\sigma$ is then derived from
\begin{equation}\label{eq:024}
    G(\sigma)= \exp[-i \sigma X] \, .
\end{equation}
In return, the algebra for any particular Lie group is obtained by differentiation
\begin{equation}\label{eq:025}
   i \frac{d}{d \sigma}G(\sigma)\Big\vert_{\sigma=0}=X_{\sigma} \, .
\end{equation}
Furthermore, we shall consider the four-dimensional spacetime manifold that has coordinates
\begin{equation} 
x_{\mu}= (t, -z, -x, -y) \, .
\label{eq:xmu311} 
\end{equation}
Lorentz transformations are
traditionally defined as the group that preserves the inner product $x^{\mu}x_{\mu}=(t^{2}-z^{2}-x^{2}-y^{2})$. 
Then we have $x_{\mu} =\eta_{\mu\nu}x^{\nu}$, where in the Minkowski metric,
$\eta_{\mu\nu} = diag(1,-1,-1,-1)$.
Specifically,
\begin{equation}\label{lgcond}
\eta_{\alpha\beta}a^{\alpha}\,_{\mu}a^{\beta}\,_{\nu}=\eta_{\mu\nu} \, ,
\end{equation}
where $a^{\alpha}\,_{\mu}$ are taken to be the components of the transformation matrix. 
Here, the reordering of coordinates  as in Equation~\ref{eq:xmu311} is convenient for our purposes, 
as in many cases one can ignore the $x,y$ coordinates and only work with 
the $t,z$ coordinates. The coordinate $z$ is distinctive as it is common practice to boost along that direction 
in special relativity. Also, in optics, the optical axis is usually chosen to be the $z$-axis.
When the Lorentz transformations are restricted to the condition that $\det a^{\alpha}\,_{\mu} = 1$, i.e., no space reflections, and to $a^{0}\,_{0} \geq 0$, i.e., \mbox{no time inversion}, this group is called the proper, orthochronous Lorentz group $SO(3\, ,1)$.

In Section~\ref{sec:2} classical and quantum phase space is defined. In Section~\ref{sec:6} linear canonical transformations are discussed along with some symmetry groups. Section~\ref{sec:3} defines the Wigner function and the density matrix, while Section~\ref{sec:4} deals with harmonic oscillators in phase space. Section~\ref{sec:9} consists of coherent states, the squeezed vacuum, and one-mode squeezed states. In Section~\ref{sec:68} the symmetries of two-mode states and Wigner functions related to the symmetries of two-mode states are presented. In Section~\ref{sec:92} the overlap of Wigner functions and squeezed states of light are given. Section~\ref{sec:sqz06} discusses coupled harmonic oscillators and entanglement. In Section~\ref{sec:14.3} we present the density matrix and the Poincaré sphere. Then follows Section~\ref{sec:conc} which gives conclusions.
Appendix A defines the proper, orthochronous Lorentz group $SO(3\, ,1)$ and the special linear group $SL(2\, ,C)$. Appendix B details the correspondence between the annihilation and creation and the phase-space operators and gives the commutation relations for the Lorentz group $SO(3\, ,2)$.  

\section{Classical and Quantum Phase Space }\label{sec:2}

The Lagrangian
and Hamiltonian formalisms serve as
reformulations of Newton's second law in classical mechanics.
Suppose a dynamical system has $n$ independent coordinates $x_{1}, x_{2}, \cdots ,
x_{n}$ the Lagrangian
is a function not only of these coordinates and their time derivatives, but also of the time variable:
\begin{equation}\label{eq:lag1}
\mathcal{L} = \mathcal{L}(x_{1}, x_{2}, \cdots , x_{n}; \dot{x}_{1}, \dot{x}_{2}, \cdots , \dot{x}_{n};
t) \, .
\end{equation}
The momentum variable conjugate to $x$ is defined as 
\begin{equation}\label{eq:lag2}
p_{i} = \frac{\partial \mathcal{L}}{\partial \dot{x}_{i} } \, .
\end{equation} 
The Hamiltonian is then defined as 
\begin{equation}\label{eq:ham14} 
\mathcal{H} = \sum_{i} \dot{x}_{i} p_{i}  - \mathcal{L} \, .
\end{equation}
For each $i$, the Hamiltonian equation of motion can be written as 
\begin{equation}\label{eq:ham16}
\dot{x}_{i} = \frac{ \partial \mathcal{H}}{\partial p_{i} } , \qquad  \dot{p}_{i} =
-\frac{ \partial \mathcal{H}}{\partial x_{i} } \, . 
\end{equation}
The Hamiltonian can now be regarded as a function of $x_{1},
x_{2},\cdots , x_{n}$ and $p_{1}, p_{2}, $ $\cdots , p_{n}$.
Therefore, the Hamiltonian formalism describes a dynamical
system of $n$ degrees of freedom given by $n$ coordinate variables $x_{1},
x_{2}, \cdots, x_{n}$, and their conjugate momenta $p_{1}, p_{2},
\cdots , p_{n}$. Now the 
$2n$-dimensional space that is spanned by $n$ coordinate and $n$ 
momentum variables is called phase space.

For one degree of freedom, the dynamical system
can be completely determined by
a two-dimensional phase space.
For a one-dimensional harmonic oscillator, the Hamiltonian is the total energy, and the trajectory is an ellipse in the phase space of $x$ and $p$.
If there is one free particle, this corresponds to one point in phase space and traces a trajectory that is in
a line parallel to the $x$ axis, with a fixed value of $p$.
For $N$ particles, there will be $N$ points and $N$ trajectories. For
$N$ large, the problem is treated statistically.  
    
Consider a volume element in phase space, $\Delta x_{1}\Delta
x_{2} \cdots \Delta x_{n} \Delta p_{1} \Delta p_{2} \cdots
\Delta p_{n}$. This then has a probability distribution
function given by $f(x_{1} ,x_{2}, \cdots ,x_{n} ,p_{1} ,p_{2} , \cdots , p_{n} ,t)$
such that the distribution function is normalized:
\begin{equation} 
\int f(x_{1} ,x_{2}, \cdots ,x_{n} ,p_{1} ,p_{2} , \cdots ,
p_{n} ,t)~dx_{1}dx_{2} \cdots d x_{n} dp_{1} dp_{2} \cdots dp_{n} = 1 \, ,
\label{eq:171} 
\end{equation}
because the total number of particles $N$ is distributed throughout the phase space.
If our interest is in distributions in one pair of 
variables, $x_{1}$ and $p_{1}$, then the probability distribution in
this two-dimensional phase space is:
\begin{equation} 
f(x_{1} ,p_{1} ,t) 
= \int f(x_{1} ,x_{2}, \cdots ,x_{n} ,p_{1} ,p_{2} , \cdots ,
p_{n} ,t) \, dx_{2} \cdots d x_{n} dp_{2} \cdots dp_{n} \, . 
\label{eq:172} 
\end{equation} 

The phase space distribution
function for a given $f(x_{1} ,p_{1} ,t)$ does not
have to be unique~\cite{fey72}, as
a function $g(x_{1} ,p_{1} ,t)$ can always be constructed such that 
\begin{equation} 
\int g(x_{1}, p_{1}, t) dx_{1} dp_{1} = 0  \, ,
\label{eq:174} 
\end{equation} 
can be added to $f(x_{1} ,p_{1} ,t)$.  This then enables us to
define a quantum phase space drastically different from a classical
phase space.

The time derivative of the distribution function $f(x_{1} ,p_{1} ,t)$ is given by 
\begin{equation} 
\frac{ df}{dt} = \sum _{i} \left ( \frac{ \partial f}{\partial x_{i}
}\dot{ x}_{i} + \frac{ \partial f}{\partial p_{i} } \dot{ p}_{i}
\right ) + \frac{ \partial f }{ \partial t} \, .
\label{eq:175} 
\end{equation} 
Consider writing the time derivative of $f$ given in 
Equation~(\ref{eq:175}) as 
\begin{equation} 
\frac{ df }{dt} = \sum _{i} \left ( \frac{ \partial }{ \partial x _{i}
}  \left ( f \dot{ 
x}_{i} \right ) + \frac{ \partial   } {\partial p_{i} } \left ( f
\dot{ p}_{i}  \right ) \right ) + \frac{ \partial f}{
\partial t } \, .
\label{eq:178}
\end{equation} 
This is equivalent to Liouville's continuity equation in phase space.

\section{Linear Canonical Transformations in Quantum Mechanics}\label{sec:6}

The transformations which leave the form of Hamilton's equations 
invariant are called canonical transformations in classical mechanics~\cite{goldstein80}. 
The transformation from one-pair of variables ($x,p$) to ($x',p'$)
is canonical when
\begin{equation} 
\frac{ \partial x'}{ \partial x} \frac{ \partial p' }{ \partial p } - \frac{\partial p'}{ \partial x} \frac{ \partial x'}{ \partial p } = 1 \, .\label{eq:41} 
\end{equation} 
Thus in two-dimensional phase space, canonical transformations are
those that are area-preserving. The group of homogeneous linear canonical transformations consists of rotations and squeezes and is called $Sp(2)$~\cite{kiwi90b}. When  translations are added, the group is inhomogeneous, then we use the augmented coordinate system of $(x,p)$ as $(1,x,p)$. 
The transformations of this group acting on $(1,x,p)^T$ can be represented by a $3\times3$ matrix:
\begin{equation} \label{eq:43a}
\begin{pmatrix} 1 & 0 \cr v  &  M\end{pmatrix}
\, , 
\end{equation}
where $M$ belongs to $Sp(2)$ and $v=(a,b)^T$ with $a,b$ real parameters. This group is called $ISp(2)$.

Let us first consider translations in phase space that take the form
\begin{equation} 
x' = x + a ,  \qquad  p' = p + b \, .\label{eq:42} 
\end{equation} 
These can be represented by defining the translation matrix $T(a,b)$ and writing
\begin{equation} 
\begin{pmatrix} 1 \cr  x^{\prime} \cr p^{\prime}  \end{pmatrix}  =
\begin{pmatrix} 1 & 0  & 0 \cr a   &1  &  0\cr b &0 & 1 \end{pmatrix}
\begin{pmatrix} 1\cr  x \cr p \end{pmatrix}  
\, . \label{eq:43} 
\end{equation} 
Two translation matrices $T(a,b)$ and $T(a',b')$ commute with each other, and
hence the translations form an invariant group. The matrices are also commutative, so these matrices form an Abelian invariant group.
This translation group will now be supplemented by
the rotation around the origin by $(\theta /2)$ that has the augmented form
\begin{equation} 
R(\theta ) = \begin{pmatrix} 1 & 0 & 0  \cr 0 &  \cos( \theta /2)   & - \sin (\theta /2)  \cr 0  & \sin (\theta /2) & \;\;\cos (\theta /2)    \end{pmatrix} \, . \label{eq:45} 
\end{equation} 
Rotations form a one-parameter group.  This rotation matrix can be
multiplied by the translation matrix of Equation~(\ref{eq:43}) or vice versa, but the two
matrices do not commute with each other as can be easily seen by direct multiplication.  

The transformations consisting of rotations around the one axis and translations in two dimensions
is called the two-dimensional Euclidean group, known as $E(2)$.
This group occupies an important place in the theory of massless particles~\cite{wig39,bknp24} and is the fundamental language for the coherent state representation of light~\cite{perelomov86}. 

Let us consider the generators of the transformation given in Equations~(\ref{eq:43}) and~(\ref{eq:45}). The translation matrix $T(a,b)$ given in Equation~(\ref{eq:43}) 
can be written as 
\begin{equation} 
T(a,b) = e^{-i \left ( aP_{1} + bP_{2} \right ) } \, , \label{eq:49} 
\end{equation} 
where the generators, $P_1$ and $P_2$, are 
\begin{equation}\label{eq:p1p2}
P_{1} = \begin{pmatrix}  0&0&0 \cr i & 0 & 0  \cr 0 & 0
& 0 \end{pmatrix} , \quad P_{2} =  \begin{pmatrix} 0
& 0 & 0 \cr 0 & 0 & 0 \cr i & 0 & 0  \end{pmatrix} \, .
\end{equation}

The rotation matrix is generated by
\begin{equation} 
J_2 = \begin{pmatrix}  0 & 0 & \;\;\;0  \cr  0  & 0 & -i/2
\cr 0   & i/2 & \;\;\;0
\end{pmatrix} \, , \label{eq:411} 
\end{equation} 
and
\begin{equation} 
R(\theta ) = e ^{-i \theta J_2} \, . \label{eq:412} 
\end{equation} 
The generators  $J_2$, $P_{1}$, and $P_{2}$  satisfy the following commutation relations:
\begin{equation}
[P_{1} , P_{2} ] = 0 \, ,\qquad [P_{1} , J_2] = -\frac{i}{2}P_{2} \, , \qquad  [P_{2} , J_2] = \frac{i}{2} P_{1} \, .\label{eq:413} 
\end{equation}
Indeed, $J_2$, $P_{1}$, and $P_{2}$ satisfy closed commutation
relations forming the Lie algebra for
the two-dimensional Euclidean group consisting
of rotations and translations in a two-dimensional space~\cite{wig39, hks82}.

In addition to rotations and translations, phase space also allows squeezes 
$S(\theta,\lambda)$ where $\lambda$ is the hyperbolic squeeze parameter which describes the strength of the squeeze.
We have $e^{\lambda} > 0 \quad \mbox{and} \quad e^{-\lambda} > 0 \quad \mbox{for all real } \lambda.$
The range of the angle $\theta$ 
is $0 \leq \theta < 2\pi$. Both $\lambda$ and $\theta$ are dimensionless parameters.  Then $S(0,\lambda)$ takes the form 
\begin{equation} 
S(0,\lambda) = \begin{pmatrix} 1 & 0 &0 \cr 0 & e^{\lambda /2} & 0 & \cr 0 & 0 
& e^{-\lambda /2}\end{pmatrix}  \label{eq:48} 
\end{equation} 
for the augmented coordinate system as in Equation~(\ref{eq:43a}).
The elongation along the $x$ axis necessarily yields to the
contraction along the $p$ axis to preserve the area in the phase space. For the squeeze along $(\theta /2)$ direction we use $R(\theta)$ from Equation~(\ref{eq:45}), so
 we have 
\begin{eqnarray}\label{eq:139} 
& &  R(\theta)S(0,\lambda)R(-\theta )\nonumber \\[2mm]
& & \hspace{5mm} =  \begin{pmatrix} 1& 0 &0 \cr 0&  \cosh (\lambda /2) + ( \cos \theta )\sinh (\lambda /2 ) & (\sin \theta)
\sinh (\lambda /2)   \cr 0 &  (\sin \theta)\sinh (\lambda /2) & \cosh (\lambda
/2) - (\cos \theta )\sinh (\lambda /2)  \end{pmatrix} \, .
\end{eqnarray} 
The matrix in Equation~(\ref{eq:48}), written in two-dimensional form, satisfies the symplectic condition~\cite{bknp24}:
\begin{equation} 
S(\theta ,\lambda ) J S( \theta ,\lambda )^T = J \, ,\label{eq:140} 
\end{equation}
where $J$ is
\begin{equation} 
J =  \begin{pmatrix}
  0 & 1 \cr -1 & 0  \end{pmatrix} \, .\label{eq:131} 
\end{equation}
 
Since a canonical transformation in quantum mechanics can be
followed by another canonical transformation, the most general
form of the transformation matrix is a product of the above
three forms of matrices. This mathematics is simplified by
using the generators of the transformation matrices.  

The squeeze matrix of Equation~(\ref{eq:48}) can be written as:
\begin{equation} 
S(0,\lambda) = e^{-i\lambda K_3} \, . \label{eq:414} 
\end{equation}  
The generator $K_3$ is given in Table \ref{tabl1}. Its augmented  form is: 
\begin{equation}
\begin{pmatrix} 1 & 0 & 0 \cr 0 &i/2 & 0 \cr 0 & 0 & -i/2
\end{pmatrix} \, . \label{eq:4139a}
\end{equation} 

Since the squeeze matrix of Equation~(\ref{eq:48}) generates symplectic
transformations, we consider first the 
symplectic group $Sp(2)$~\cite{guil84,perelomov86} which is also known 
as $SL(2\, ,R)$ the special real linear group in two-dimensions. As the Lorentz
transformations associated with $SL(2\, ,R)$ are real, the generators must be imaginary. 
The following generators form the Lie algebra of the $Sp(2)$ group.
They are the rotation generator $J_2$ and the boost generators $K_1$, and
$K_3$ as given in two-by-two form in Table~(\ref{tabl1}).
The commutation relations for these generators are 
\begin{equation} 
[K_{1} \, , K_{3} ] = -iJ_2 , \qquad  [K_{3}\, , J_2] = -iK_{1} , \qquad [K_{1} , J_2] = iK_{3} \, .\label{eq:416a} 
\end{equation}
The group generated by the three generators in Equation~(\ref{eq:416a}) operates in the
$(x,~p)$ plane and is
isomorphic (one-to-one and onto) the Lorentz group $SO(2\, ,1)$ which operates in the
$(x,~y)$ plane. The connection between the two groups has been extensively discussed
in the literature~\cite{bknp24}. 

However, the generators for the Lorentz group $SO(2\, ,1)$ are different.
They are  $K_1$, $K_2$, and $J_3$.
The matrices $J_3$, $K_{1}$, and $K_{2}$
as given in Table~(\ref{tabl1}) satisfy
the following commutation relations:
\begin{equation} 
[K_{1} , K_{2} ] = -iJ_3 , \qquad  [K_{1} , J_3] = -iK_{2} , \qquad  [K_{2} , J_3] =
iK_{1} .\label{eq:416} 
\end{equation} 
This set of commutation relations forms the Lie algebra for
the $SO(2\, ,1)$ Lorentz group~\cite{bknp24} that acts mainly in the $(x,~y)$ plane. 
Although this Lorentz group is applicable to the $(t, x, y)$ coordinates, rotation
around the $z$-axis which will not change the group,
will extend the representation to the full $(t, z, x, y)$
space. Calculations in high-energy physics that involve
Lorentz transformations are frequently based on
$SO(2\, ,1)$~\cite{bknp24}. This group has also become a basic
language in classical and quantum optics~\cite{gerry_01,wunsche_2000}.

The inhomogeneous symplectic group in the
two dimensions $ISp(2)$ or $ISL(2\, R)$ has also found some applications
in canonical transformations~\cite{kramer82,ellinas2019}. This group
is formed by adding a translation component to the $Sp(2)$ group.
If we take into account the translation operators, the commutation
relations become
\begin{eqnarray} \label{eq:417} 
& &[K_{1}, P_{1}] = \frac{i}{2}  P_{2}, \qquad  [K_{1},P_{2} ]
= \frac{ i}{2} P_{1} \nonumber \\[1.0ex]
& &[K_{3}, P_{1}]  =  \frac{i}{2} P_{1},  \qquad [K_{3}, 
P_{2}] = - \frac{i}{2} P_{2} \, .
\end{eqnarray} 
These commutators together with those of Equation~(\ref{eq:416a}) form the set of closed commutation relations (or Lie algebra) for the group of canonical transformations in quantum mechanics.
This group is the inhomogeneous symplectic group in the
two dimensions or $ISp(2)$.

\begin{figure}[!ht]
\begin{center}
\begin{tikzpicture} 

 
\draw[black,  thick, ->, >=stealth] (0.0,3) -- (6,3);
\draw[black, thick,->, >=stealth] (3, 1) -- (3, 5); 
\draw[black,  thick,  ->, >=stealth] (9, 3) -- (13, 3);
\draw[black,  thick, ->, >=stealth] (11, 1) -- (11, 5); 
\draw[black, thick, ->, >=stealth] (9, 1) -- (13, 5);

\draw[ultra thick, blue] (3,3) circle (1cm);
\draw[ultra thick,green] (3,3) ellipse (2.5 cm and 0.4 cm);
\draw[ultra thick, blue] (11,3) circle (1cm);
\draw[ultra thick,red, rotate around={45.0:(11,3)}] (11,3) ellipse (2.5 cm and 0.4 cm);
 

\path (6.2,3) node [black, scale=1.1] {$x$};
\path (3,5.2) node [black, scale=1.1] {$p$};
\path (13.2,3) node [black, scale=1.1] {$x$};
\path (11,5.2) node [black, scale=1.1] {$p$};

\path (3, 0.5) node [black] {Squeeze $\Leftrightarrow$ Boost along $z$};
\path (11,0.5) node [black] {Squeeze+Rotation $\Leftrightarrow$ Boost along $x$};

\end{tikzpicture}
\end{center}
\caption{The transformations of $Sp(2)$ produce rotations and
squeezes. The circle in this figure corresponds to a
Gaussian distribution. As shown on the left, the Lorentz boost of $SO(2\, ,1)$ along 
the $z$-direction corresponds to the squeeze of $Sp(2)$  along the x-axis of the phase space. On the right is the action of the rotated squeeze of $Sp(2)$, which transforms the circle into a tilted ellipse whose major axis is aligned along the $p=x$ line.  This transformation corresponds to the boost along the $x$-direction of $SO(2\, ,1)$.}
\label{fig:sp00}
\end{figure}
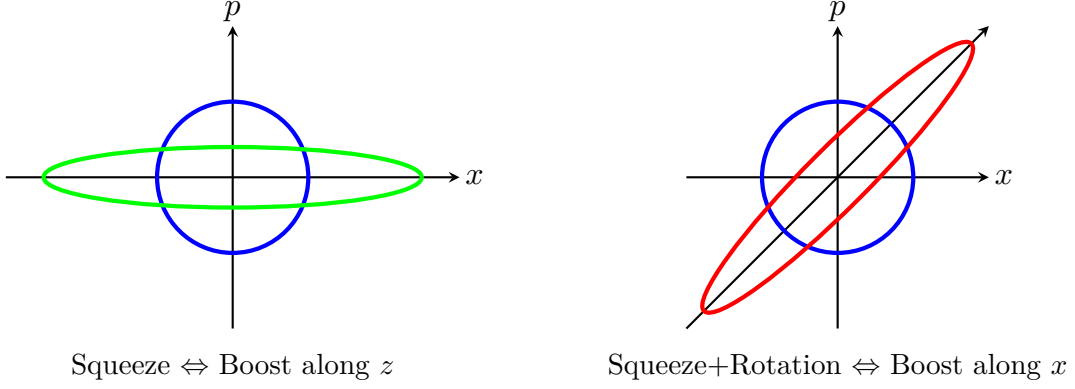

In Figure~\ref{fig:sp00} we illustrate the $Sp(2)$ group. Lorentz transformations in the Wigner phase space are symplectic transformations. 

\FloatBarrier

\section{The Wigner Function and the Density Matrix}\label{sec:3}
Consider a wave function that depends on $x_{1},x_{2},\cdots, x_{n}$, and
$t$.  We can define the density matrix for this wave function as 
\begin{equation} 
\rho (x_{1}, \cdots ,x_{n} ;x'_{1}, \cdots ,x'_{n};t) = \sum _{k}
w_{k}  \psi_{k} (x_{1} , \cdots ,x_{n} ,t) \psi _{k} ^{*} (x' _{1},
\cdots , x'_{n},t) \, . \label{eq:31} 
\end{equation} 
Then it is straight-forward to generalize the single
variable-density matrix~\cite{blum_2012,neum2018}, with the normalization constant
\begin{equation} 
\sum _{n} w_{k}  = 1 . \label{eq:32} 
\end{equation} 
The density matrix forms the transition from the 
Schrödinger wave function to the Wigner function. 
For $n$ pairs of canonical variables, the Wigner
function can be written as~\cite{wig32a,wig87} 
\begin{eqnarray} 
& & W \left ( x_{1} , \cdots ,x_{n} ;p_{1} ,\cdots ,p_{n};t
\right ) = 
\left ( \frac{1}{ \pi} \right ) ^{n}  \int  \exp \left \{ 2i \left (
p_{1} y_{1} + \cdots  + p_{n} y_{n} \right ) \right \}  \nonumber \\[2ex]
& & \hspace{4mm}\times \rho \left (  x_{1} - y_{1} ,\cdots
,x_{n} - y_{n} ; x_{1} + y_{1} 
,\cdots ,x_{n} + y_{n}  ;t \right )\, dy_{1} dy_{2} \cdots dy_{n}  \, .
\label{eq:33} 
\end{eqnarray} 
Here, $(1/\pi)$ forms the normalization constant,  $x_{i}$ and $p_{i}$ are c-numbers, and the
Wigner function is defined over the
$2n$-dimensional phase space.  When the system is in a pure state with the wave
function $ \psi (x_{1} ,\cdots ,x_{n} ,t)$, we have
\begin{eqnarray} 
& & W(x{_1} ,\cdots ,x_{n} ;p_{1} ,\cdots ,p_{n} ;t) =  \left (
\frac{1}{\pi} \right ) ^{n} \int  \exp \left \{ 2i \left ( p_{1} y_{1} + \cdots
+ p_{n} y_{n} \right ) \right \}  \nonumber \\[2ex]
& & \hspace{4mm} \times \psi ^{*} (x_{1} + y_{1} ,\cdots ,x_{n} + y_{n} ,t) \psi
(x_{1} - y_{1} ,\cdots ,x_{n} - y_{n} ,t) \,dy_{1} \, dy_{2} \cdots \, dy_{n}
\, .\label{eq:34} 
\end{eqnarray} 
To study the properties of the pure-state
Wigner function let us
start with the simplest form, which is when the system depends only on
one pair of $x$ and $p$ variables.
Then the Wigner function takes the form
\begin{equation} 
W(x,p,t) = \frac{1}{\pi} \int  e^{2ipy} \psi ^{*} (x + y,t) \psi (x -
y,t)\, ~dy \, . \label{eq:35} 
\end{equation} 
The Wigner function does not depend on time if the wave function $\psi (x)$
does not depend on time. 
Therefore, the most frequently seen Wigner function in the literature is of the form 
\begin{equation} 
W(x,p) = \frac{1}{\pi} \int  e^{2ipy} \psi ^{*} (x + y) \psi (x - y) \,
~dy \, . \label{eq:36} 
\end{equation}

The time-dependent Wigner function can be derived from a time-dependent 
wave function by using the time-dependent Schrödinger equation:
\begin{equation} 
i \frac { \partial }{ \partial t } \psi (x,t) = -\frac{1}{ 2m
}  \frac {\partial^{2}}{\partial x^{2}}   \psi
(x,t) + V(x) \psi (x,t) \, . \label{eq:315} 
\end{equation} 
Here, $m$ is the mass of the particle and $V(x)$ is the potential. Depending on the
form of $V(x)$ this can be an infinite order differential equation.  On the other hand if
\begin{equation} 
\frac { \partial ^{3}}{ \partial x^{3}} \,  V(x) = 0 \, ,
\label{eq:320} 
\end{equation} 
then Equation~(\ref{eq:315})
can be reduced to the classical Liouville equation:
\begin{equation} 
\frac{ \partial}{\partial t} W(x,p,t) =  \left (\frac {\partial \mathcal{H}}{
\partial x} \right )  \frac { \partial }{\partial p } W(x,p,t) -
\left ( \frac {\partial \mathcal{H}}{\partial p } \right ) \frac {\partial }{
\partial x} W(x,p,t) \, . \label{eq:321} 
\end{equation} 
The Wigner distribution function satisfies Equation~(\ref{eq:321}) if the potential is of the form $V(x) = ax^2 + bx +c$.
Although the classical phase-space and the Wigner distribution functions share the same mathematics,
the classical phase-space distribution is a probability 
distribution in phase space~\cite{fey72,goldstein80}, while the Wigner
distribution is not~\cite{wig32a,wig87}. 

Let us consider that a system is in the state $\psi (x)$. When an observation is made
that results in the state vector of the system becoming $\phi (x)$, the probability
of this observation is $\mid (\psi ,\phi ) \mid ^{2}$.
This is the absolute square
of the scalar product of the two state vectors.  If the Wigner function is used, 
the transition probability between the two states takes the form 
\begin{equation} 
\mid (\psi(x) ,\phi (x) )\mid ^{2}   = 2\pi \int W_{\psi} (x,p) W_{\phi}
(x,p) ~dx~dp . \label{eq:312} 
\end{equation} 
If $\psi$ and $\phi$ are orthogonal,
Equation~(\ref{eq:312}) must vanish everywhere in phase space.
This is not possible if both Wigner functions are
positive everywhere in phase space and thus provides clear proof that
the Wigner function cannot be positive everywhere in phase space.

\begin{figure}[!ht]
\centering
\begin{tikzpicture} [align=center]


\draw[thick, green, dashed] (-4.3, 1.0) -- (4.3, 1.0);

\draw[ultra thick, rotate around={14.0:(0,0)},red] (0,0) ellipse (4cm and 0.25cm);
\draw[ultra thick, blue] (0,0) circle (1cm);

\draw[black, ->, >=stealth] (-4.3, 0) -- (4.3, 0); 
\draw[black,  ->, >=stealth] (0, -1.3) -- (0, 2.3);
\path (4.5,0) node [black, scale=1.1] {$x$ };
\path (0,2.5) node [black, scale=1.1] {$p$ };

\draw[thick, brown, dashed] (-4.3, 1.0) -- (4.3, 1.0);
\draw[thick, brown, dashed] (-4.3, -1.0) -- (4.3, -1.0);
\draw[thick, <->, stealth-stealth, color=violet] (-3.9, -1.5) -- node (M4)[fill=white,midway, scale=1.1] {Spread} (3.9, -1.5);

\draw[thick, ->, >=stealth, color=violet] (0, 1.4) -- node (M2)[fill=white,midway, scale=1.0] {$(p/m)t$} (3.87, 1.4);

\end{tikzpicture}
\caption{The spread of the Gaussian wave packet. The circle represents the Gaussian distribution, while the tilted ellipse represents the time development~\cite{bkn19iop}. }\label{fig:shear}
\end{figure}
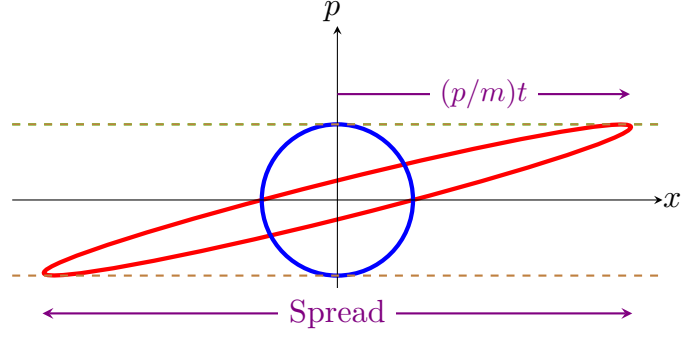

Let us explore the Wigner function for two simple values of the potential in Equation~(\ref{eq:315}): \\
i)~ If the particles are free, the potential $V(x) = 0 $, the quantum Liouville equation reduces to
\begin{equation} 
\frac { \partial }{ \partial t} W(x,p,t) = -\left ( \frac {p}{m}
\right ) \frac { \partial }{ \partial x} W(x,p,t)\, .\label{eq:3240} 
\end{equation} 
The solution of this differential equation is of the form
\begin{equation} 
W(x,p,t) = W(x - pt/m, p) \, .\label{eq:3250} 
\end{equation}
This solution is represented through a canonical coordinate transformation:
\begin{equation} 
\begin{pmatrix} 1 \cr x' \cr p'   \end{pmatrix} = 
\begin{pmatrix}  1 & 0 & 0 \cr 0 & 1 & t/m  \cr 0 & 0 & 1\end{pmatrix}
 \begin{pmatrix} 1  \cr x \cr p \end{pmatrix}  \, .
\label{eq:326a}
\end{equation} 
This corresponds to the wave packet spread. If the Wigner function 
has a Gaussian distribution of the form of
\begin{equation}\label{eq:326b}
W(x, p , t=0) =  \frac{1}{\pi} \,
\exp \left \{-\left(x^{2} + p^{2} \right) \right \} \, ,
\end{equation}
the time evolution of
the Wigner function becomes
\begin{equation} \label{eq:3271}
W(x,p,t)=\frac{1}{\pi}\exp\{{-[(x-pt/m)^{2},p^{2}]} \} \, .
\end{equation}
This distribution is concentrated around the Gaussian distribution.
A transformation of this type is also known as "shear".  This is illustrated in Figure~\ref{fig:shear}.

ii)~ If the potential is $V(x) = gx $, as in the case of a particle in a constant
gravitational field, the quantum Liouville equation becomes
\begin{equation} 
\frac { \partial }{ \partial t} W(x,p,t) = -\left ( \frac {p}{m}
\right ) \frac { \partial }{ \partial x} W(x,p,t) + g \frac { \partial
}{ \partial p } W(x,p,t) \, .\label{eq:324} 
\end{equation} 
If the Wigner distribution function is
known at $t = 0$, then the solution of
this differential equation is
\begin{equation} 
W(x,p,t) = W(x - pt/m, p + gt) \, .\label{eq:325} 
\end{equation} 
As time progresses, the phase-space distribution moves from $p = 0$ to $p = -gt$.  At the same time, $x = 0$ moves to $pt/m$.  The solution can
therefore be represented by a canonical coordinate transformation:
\begin{equation} 
\begin{pmatrix} 1\cr x' \cr p'   \end{pmatrix}  = 
\begin{pmatrix}  1 & 0 & 0 \cr 0 & 1 & t/m \cr -gt & 0 & 1 \end{pmatrix}
\begin{pmatrix} 1 \cr x \cr p  \end{pmatrix} \, .
\label{eq:326} 
\end{equation} 
This means that the solution of the Liouville equation for the linear
potential can be represented by a linear canonical coordinate transformation.

Using the time independent Wigner function as given in Equation~(\ref{eq:36}), 
it is possible to convert the
Wigner function back to the density matrix:
\begin{equation} 
\rho (x,x') = \int  W \left ( \frac {x + x'}{2}, p \right ) e
^{-ip(x-x')}~dp \, .\label{eq:355} 
\end{equation}
Using this relation, we can calculate the trace of the density function as $Tr(\rho )$ and $Tr(\rho ^{2})$ from the Wigner
function.  
From Equation~(\ref{eq:355}), $Tr(\rho )$ is
\begin{equation} 
Tr(\rho ) = \int  \rho (x,x)dx = \int  W(x,p) ~dx ~dp = 1 \, .\label{eq:356} 
\end{equation} 
Two different Wigner functions $W_{1}$ and $W_{2}$ can be constructed from
two different density matrices $\rho _{1}$ and $\rho _{2}$.  Then,
\begin{eqnarray} 
Tr\left ( \rho_{1} \rho _{2} \right ) & = & \int \left
\{ \int  \rho _{1} 
(x,x') \rho _{2} (x',x)~dx' \right \} ~dx  \nonumber \\[2mm]
& = &   2\pi \int  W_{1} (x,p) W_{2} (x,p) ~dx ~dp \, .\label{eq:357} 
\end{eqnarray} 
Consequently,
\begin{eqnarray} 
Tr(\rho ^{2} )  & = &  \int \rho (x,x') \rho (x',x) ~dx^{\prime} ~dx \nonumber \\[2mm]
 & = &  2\pi \int \left (  W(x,p) \right ) ^{2}  ~dx ~dp . \label{eq:358} 
\end{eqnarray} 
As can be seen, this quantity is equal to one.  The density matrix is in a pure state. 

The best known example of both the density function and the
corresponding Wigner function is the harmonic oscillator in thermal
equilibrium~\cite{fey72}. The density matrix is given as
\begin{equation} 
\rho _{T} (x,x') = \left ( 1 - e ^{-\omega /kT} \right )
 \sum _{n} e^{-n\omega /kT} \psi _{n} (x) \psi _{n}^{*} (x') \, .\label{eq:288} 
\end{equation} 
Here, $\omega$ is the energy separation, $T$ is the temperature, and $k$ is the Boltzmann constant.
For this matrix, $Tr(\rho )$ is 
\begin{equation} 
Tr( \rho _{T}) = \int \rho _{T} (x,x) ~dx = \left ( 1 - e^{-\omega /kT}
\right ) \sum _{n} e ^{-n\omega /kT} . \label{eq:289} 
\end{equation} 
This matrix becomes one after summation. 
Then,
$Tr \left ( \rho _{T}\,^{2} \right )$ is 
\begin{equation} 
Tr(\rho _{T} \rho _{T} ) = \int \left \{ \int \rho _{T} (x,x') \; \rho
_{T} (x',x) ~dx' \right \} ~dx . \label{eq:290}          
\end{equation} 
From the expression of Equation~(\ref{eq:288}),
\begin{equation} 
Tr \left ( \rho _{T}\,^{2} \right )  =  \left ( 1 - e^{-\omega /kT}
\right ) ^{2} \sum _{n} e ^{-2n\omega /kT} 
= \tanh( \omega / 2kT) . \label{eq:291} 
\end{equation} 
This is less than one, and becomes one only when $T$ becomes zero.  When $T = 0$, 
the system is in a pure state. This is the ground-state of the
harmonic oscillator. 

The Wigner function for this harmonic oscillator is
\begin{equation} 
W_{T} (x,p) = \left ( 1 - e ^{-1/kT} \right ) \sum _{n} e^{-n/kT}W_{n} (x,p)
\, ,\label{eq:360} 
\end{equation} 
where $W_{n} (x,p)$ is the Wigner function for
the $n^{\rm th}$ excited-state harmonic
oscillator. $W_{n} (x,p)$ will be derived in the following section.

\section{Harmonic Oscillators in phase space}\label{sec:4} 

The one-dimensional harmonic oscillator using the classical Hamiltonian is 
\begin{equation} 
\mathcal{H} = \frac {1}{2m} p^{2} + \frac{K}{2} x^{2} \, .\label{eq:339} 
\end{equation} 
The $p$ and $x$ can be measured in units of $\sqrt{m}$ and
$1/\sqrt{K}$ respectively.   
Rewriting the Hamiltonian as
\begin{equation}
\mathcal{H} = \frac{1}{2} \left (  p^{2} + x^{2} \right ) \, ,\label{eq:340} 
\end{equation} 
the Schrödinger equation in these units takes the form 
\begin{equation} 
i \frac {\partial}{\partial t}\psi (x,t) = - \frac{1}{2}
\frac{\partial ^{2}}{\partial x^{2}}  \, \psi (x,t)
+  \frac {1}{2} x^{2} \,  \psi (x,t) \, . \label{eq:341} 
\end{equation} 

For the Wigner function the Liouville equation is then 
\begin{equation} 
\frac { \partial }{ \partial t} W(x,p,t) =  \left ( x \frac {\partial
}{ \partial p } - p \frac {\partial }{\partial x }\right ) W(x,p,t)
\, .\label{eq:342} 
\end{equation} 
The solution to this equation has the form 
\begin{equation} 
W(x,p,t) = \left \{ \exp  \left [ \left ( x \frac{\partial }{\partial
p}  - p \frac {\partial}{\partial x} \right ) t \right ] \right \}
W(x,p,0) \, .\label{eq:343} 
\end{equation} 
This is a rotation in phase space around the origin, which
can be written in matrix form as
\begin{equation} 
\begin{pmatrix} x' \cr p' \end{pmatrix}     = 
\begin{pmatrix}  \cos \theta  & -\sin \theta \cr \sin \theta & \;\;\;\cos \theta
\end{pmatrix} \begin{pmatrix} x \cr p \end{pmatrix} 
\, . \label{eq:344} 
\end{equation} 
The distribution is independent of 
time when invariant under rotations.

The time-independent Schrödinger equation is 
\begin{equation} 
-\frac {1}{2} \frac {\partial ^{2}}{ \partial
x ^{2}} \,   \psi (x,t) +   \frac {1}{ 2} x^{2} \,
\psi (x,t) = (n + 1/2) \psi (x,t)   \label{eq:345} 
\end{equation} 
and has the normalized solutions
\begin{equation} 
\psi _{n} (x) = [1/( {\pi} 2^{n} n!)]^{1/2} H_{n} (x)
\exp(-x^{2} /2) \, .\label{eq:346} 
\end{equation} 
$H_{n} (x)$ is the Hermite polynomial
of the $n^{\rm th}$ order.  This wave 
function is thus in the energy eigenstate and hence the Wigner function can be evaluated as:  
\begin{equation} 
W_{n} (x,p) = \frac {1}{ \pi} \int  \psi _{n} ^{*}  (x + y) \psi_{n}
(x - y)e^{2ipy} dy \, .\label{eq:347} 
\end{equation} 
This calculation results in 
\begin{equation} 
W_{n} (x,p) = \left (  \frac {n!}{ \pi} \right ) \, \exp (-r^{2}
/2)  \sum _{k=0} ^{\infty} (-1)^{k}  r^{2(n-k)} / \left (
[(n-k)!]^{2} k! \right )  \, ,\label{eq:348} 
\end{equation} 
where
\begin{equation}
r^{2}  = 2(x^{2}  + p^{2} )  \, .
\end{equation}

Since now $W_{n} (x,p)$ is a function only of $r$, we can write 
$W_{n} (r)$.  This satisfies the differential equation~\cite{hillery84}:
\begin{equation} 
- \frac{1}{2r} \left [ \frac {d}{d \rho } r \left ( \frac
{d}{dr}W_{n} (r)\right ) \right ] + \frac {1}{2} r^{2} W{_n} (r) = (2n
+ 1) W_{n} (r) \, .\label{eq:349} 
\end{equation} 

Readily available in the literature~\cite{schleich88}, the solution to this equation 
has the form 
\begin{equation} 
W_{n} (x,p) =  \frac{(-1)^{n}}{\pi} \left [ L_{n} (r^{2}
)\right ] e ^{-r^{2} /2} \, ,\label{eq:350} 
\end{equation} 
where $L_{n} (r^{2} )$ is the Laguerre polynomial~\cite{arfken13}.  This expression, invariant 
under rotations around the origin, can be written as~\cite{schleich88} :
\begin{equation} 
W_{n} (x,p) = 
\frac {1}{ \pi} \left ( \frac{1}{4} \right ) ^{n}
e^{-(x^{2} + p ^{2} )} \sum _{k=0} ^{n} \left ( \frac{1}{ k! (n-k)!}
\right ) H_{2k} ( \sqrt{2}x) H_{2(n-k)} ( \sqrt{2}p) \, , \label{eq:351} 
\end{equation} 
where $H_{2k} ( \sqrt{2}x)$ and $H_{2(n-k)}( \sqrt{2}p)$ are the
Hermite polynomials.

\begin{figure}[t]
\begin{center}
\begin{tikzpicture}[scale=0.75, transform shape]
\def\tmax{3}  
\def\N{100}
\path[inner sep=0] node at (1.4,12) {\includegraphics[width=0.5\textwidth]{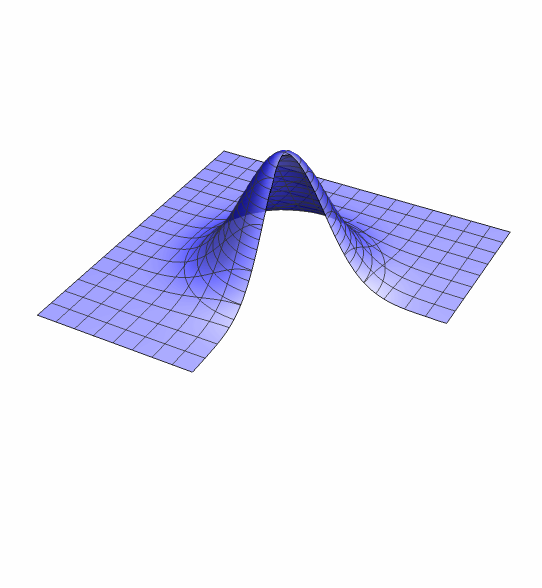}};
\path[inner sep=0] node at (10.8,12) {\includegraphics[width=0.5\textwidth]{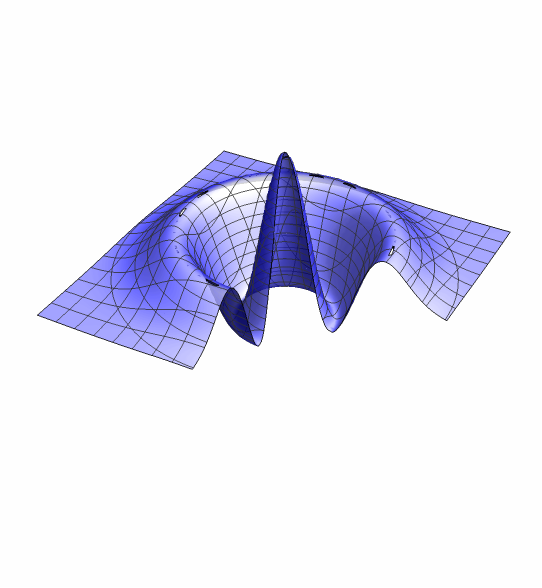}};
\path[inner sep=0] node at (10.8,5.0) {\includegraphics[width=0.5\textwidth]{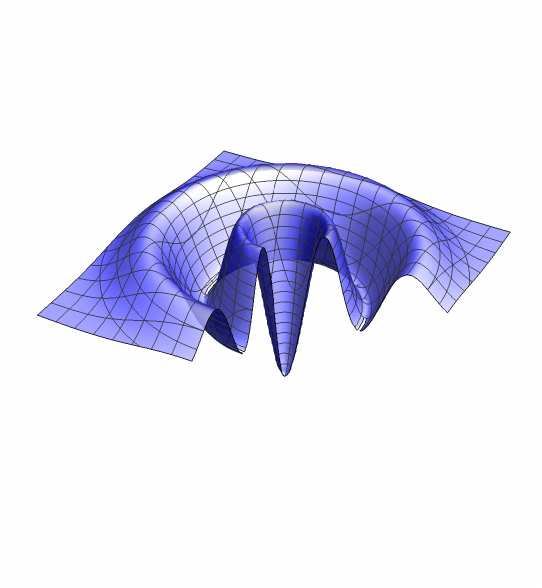}};
\path[inner sep=0] node at (1.0,5.0) {\includegraphics[width=0.5\textwidth]{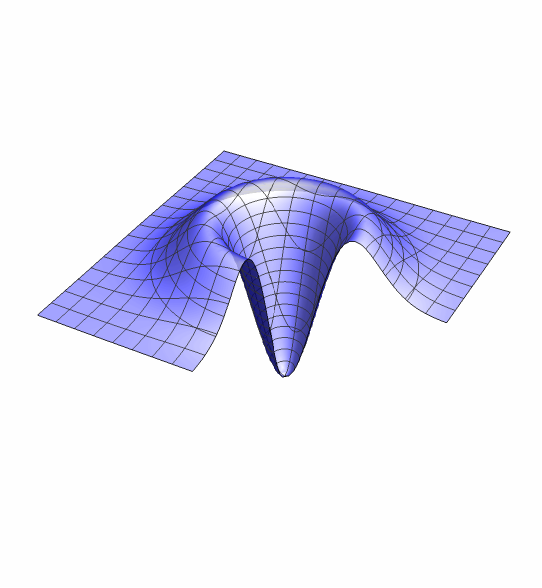}};

\path (3.4,3.5) node [black] {\Large$W_{1}(x,p)$ };
\path (12.6,3.5) node [black] {\Large$W_{3}(x,p)$ };
\path (3.0,10.6) node [black] {\Large\Large$W_0(x,p)$ };
\path(12.0,10.6) node [black] {\Large$W_{2}(x,p)$}; 
\end{tikzpicture}
\end{center}
\vspace{-2.0cm}
\caption{Wigner functions (with right-angle cross sections) of the harmonic oscillator for the ground state and for the first three excited states. The symmetry of the Wigner function with respect to $x$ and $p$ is apparent. It can also be observed that higher the excited state the function has more ripples around the origin.}\label{fig:31}
\end{figure}
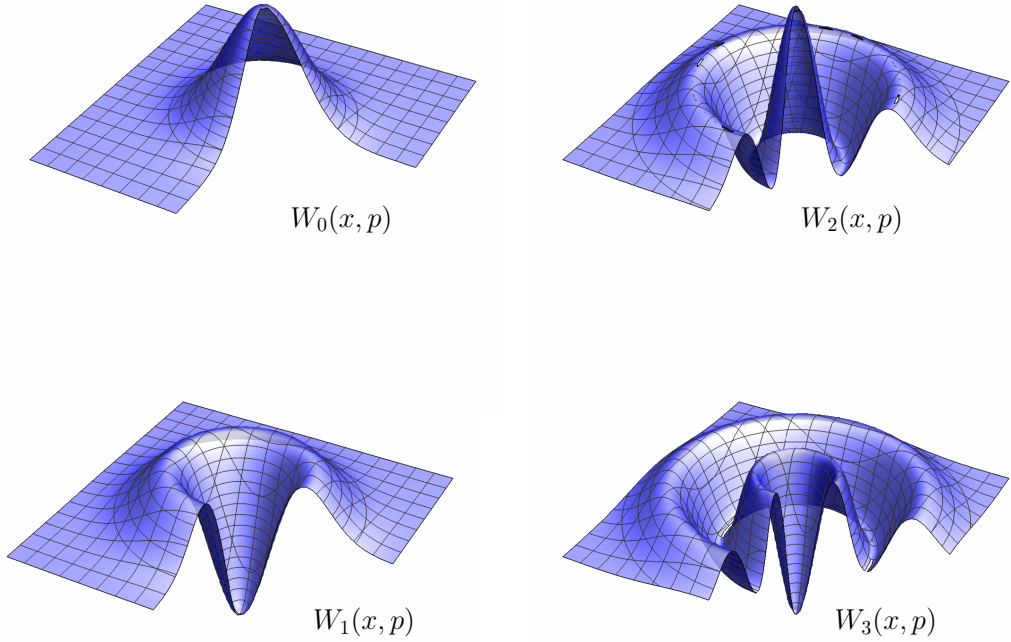

The Wigner function for the ground state is also to be drawn as the only state with no negative region:
\begin{equation} 
W_{0} (x,p) = \frac {1}{\pi} \exp \left \{ -(x^{2}  + p^{2} ) \right \} \, .\label{eq:352} 
\end{equation} 
For the first excited state, where $n = 1$,
\begin{equation} 
W_{1} (x,p) = -\frac{1}{\pi} e^{-(x^{2} + p^{2})} \left \{1-2( x^{2}  + p^{2}) \
\right \} \, . \label{eq:353} 
\end{equation} 
The second and third excited states are for $n = 2$ and $n=3$ resulting from
Equation~(\ref{eq:352}), have the forms
\begin{eqnarray}
&& W_{2}(x,p)=\frac{1}{\pi}\,e	^{-(x^2+p^2)}\left\{1 - 4(x^2+p^2 )+ 2( x^2+p^2)^2 \right\}  \, ,\\
&&W_{3}(x,p)=-\frac{1}{\pi}\,e^{-(x^2+p^2)}\left\{1 - 6(x^2+p^2 )+6( x^2+p^2)^2 -\frac{4}{3}( x^2+p^2)^3 \right\} \, ,
\end{eqnarray} 
respectively. These Wigner functions are illustrated in Figure~\ref{fig:31}.

\begin{figure}[!t]
\centering
\begin{tikzpicture}[scale=0.75, transform shape]
\def\tmax{3}  
\def\N{100}
\path[inner sep=0] node at (0.5,2) {\includegraphics[width=0.6\textwidth]{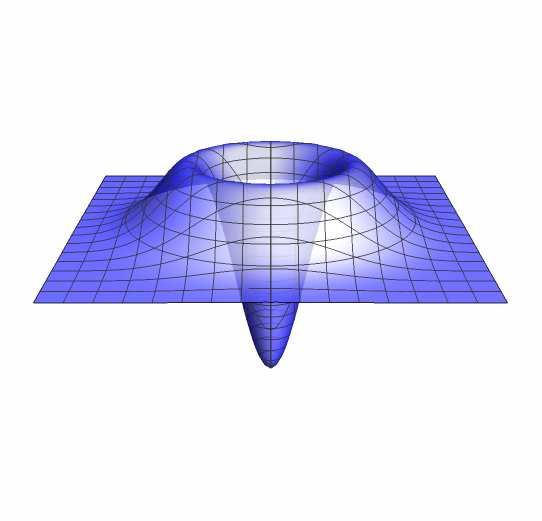}};
\path[inner sep=0] node at (10,2) {\includegraphics[width=0.7\textwidth]{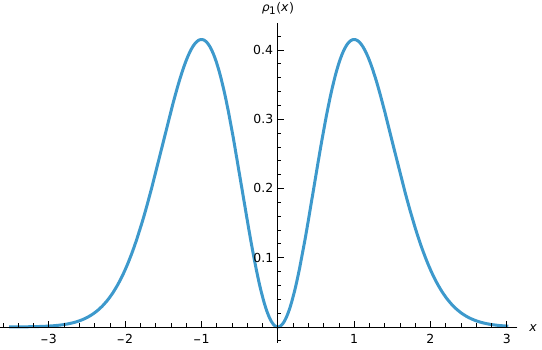}};
\path (0.5,-0.6) node [black] {\Large $W_{1}(x,p)$};
\end{tikzpicture}
\caption[The Wigner function for the harmonic oscillator.]
{The Wigner function for the harmonic oscillator in the  
first excited state.  It is negative at the origin, becomes 
positive as \mbox{\small$(x^{2} + p^{2} )$} increases, and becomes
vanishingly small as
\mbox{\small$(x^{2} + p^{2} )$} becomes very large.  If this Wigner
function is integrated  
over \mbox{\small$p$}, then it becomes the quantum probability distribution in
\mbox{\small$x$}, which is $\rho_{1}(x)=2 x^2e^{-x^2} /\sqrt{\pi }$, and is
positive for all values of \mbox{\small$x$}.} 
\label{fig:32}
\end{figure}

$W_{0} (x,p)$ is positive everywhere in phase space.  We see that $W_{1} (x,p)$ is negative at the origin, but is positive for sufficiently
large values of $(x^{2} + p^{2})$.  Both become vanishingly small for very
large values of $(x^{2} + p^{2})$.  Therefore, the probability density in
$x$ is always positive, although $W_{1}(x,p)$ is
negative around the origin. This is
illustrated in Figure~\ref{fig:32}.

In~\cite{Bell_1986} S.L. Bell discusses the relation between the negativity of the Wigner function and the issue of non-locality. It is claimed that, it is the negativity of the Wigner function that represents the true quantum nature of this formalism~\cite{mallick_2025}. Thus such a property is to be recognized as an advantage rather than an impediment~\cite{hillery84}.

The Wigner function for the
one-dimensional harmonic oscillator has been discussed.
In this coordinate system, it was shown that all
excited-state Wigner functions are invariant under rotations. Now
our interest is in canonical transformations in quantum mechanics
beginning in those states that are invariant under rotation.
Thus we start by making canonical transformations in phase space of the unit circle centered around the origin:
\begin{equation} 
x^{2}  + p^{2}  = 1 .\label{eq:439} 
\end{equation}
If the center is at $(x,p) = (a,b)$, the equation is  
\begin{equation} 
(x - a)^{2}  + (p - b)^{2}  = 1 .\label{eq:440} 
\end{equation} 
This is then a translation of the circle in Equation~(\ref{eq:439}).
Now we drop the translation from our 
consideration, and use the generators of the
symplectic group, $K_1$, $K_3$, $J_2$ as given in Table~(\ref{tabl1}) in 
two-by-two matrix form.
These generators satisfy the commutation relations of 
Equation~(\ref{eq:416a}).

\FloatBarrier
It is worthwhile to note here that $e^{-i\lambda K_{1}}$ as given in Table~\ref{tabl1} has all hyperbolic functions, while $e^{-i\lambda K_{3}}$  squeezes objects in the $xp$-plane elongating/shrinking in the $x$- and $p$-directions, respectively.
From these generators, we can construct the squeeze and rotation matrices:
\begin{equation} 
R(\theta) = e^{-i\theta J_2} \, ,\quad S(0,\lambda ) = e^{-i\lambda K_{1}}  \, \, , \quad  S(\pi /2 ,\lambda) = e^{-i\lambda K_{3} } .\label{eq:442} 
\end{equation}  

The squeeze along the $(\theta /2)$ direction was given in 
Equation~(\ref{eq:139}).
If we perform the rotation $R(\theta )$ on the circle of
Equation~(\ref{eq:439}) centered
around the origin, it remains invariant.  If the same rotation is applied
to the circle of Equation~(\ref{eq:440}), which is not centered around the origin, its effect is
\begin{equation} 
(x - a')^{2}  + (p - b')^{2}  = 1 ,\label{eq:444} 
\end{equation} 
where
\begin{equation}\label{eq:444a}
\begin{pmatrix} a' \cr b' \end{pmatrix} = 
\begin{pmatrix} \;\;\;\cos( \theta /2) & \sin( \theta /2) \cr -\sin (\theta
/2) & \cos (\theta /2) \end{pmatrix} \begin{pmatrix} a
\cr b \end{pmatrix} .
\end{equation}
This is a reflection of the transformation property given in
Equation~(\ref{eq:42}).

Under the transformation of $S(\theta ,\lambda)$, the circle of
Equation~(\ref{eq:439}) becomes a tilted ellipse:
\begin{equation} 
e^{-\lambda} \left ( x \cos \frac{ \theta }{ 2} + p \sin \frac{ \theta
}{2} \right ) ^{2} + e^{\lambda}\left( -x \sin \frac{ \theta }{2} + p
\cos \frac{\theta }{2} \right ) ^{2} = 1 \label{eq:445} 
\end{equation} 
and the circle of Equation~(\ref{eq:440}) becomes a tilted and displaced ellipse:
\begin{equation} 
\hspace{30mm}  e^{-\lambda}\left( (x - a^{\prime\prime})\cos \frac{\theta }{2} + (p - b^{\prime\prime})\sin
\frac{\theta }{2} \right ) ^{2}
+ e^{\lambda} \left ( -(x - a^{\prime\prime})\sin \frac{\theta }{2} + (p - b^{\prime\prime})\cos \frac{ \theta }{2} \right )^{2}   = 1 \, , \label{eq:446} 
\end{equation} 
with
\begin{equation} \begin{pmatrix} a^{\prime\prime} \cr b^{\prime\prime} \end{pmatrix} = 
\begin{pmatrix}  \cosh \frac{\displaystyle{\lambda}}{\displaystyle{2}} + (\cos \theta )\sinh
\frac{\displaystyle{\lambda}}{\displaystyle{2}} & (\sin \theta ) \sinh
\frac{\displaystyle{\lambda}}{\displaystyle{2}} \cr
(\sin \theta ) \sinh \frac{\displaystyle{\lambda}}{\displaystyle{2}} & \cosh \frac{\displaystyle{\lambda}}{\displaystyle{2}} -
(\cos \theta ) \sinh \frac{\displaystyle{\lambda}}{\displaystyle{2}} \end{pmatrix} 
\begin{pmatrix} a \cr b \end{pmatrix} \, . \label{eq:446a}
\end{equation}
These are area-preserving transformations.  Equation~(\ref{eq:445}) is a 
special case of Equation~(\ref{eq:446}). 
The canonical transformation in
quantum mechanics of the ground-state harmonic oscillator
provides the basic mathematical language for coherent
and squeezed states of light.  

\section{Coherent and Squeezed States of Light in Wigner's Phase Space}\label{sec:9}

The concept of free photons is well established in terms of their
creation and annihilation operators which can be defined in terms of mathematical descriptions of the energy levels of the harmonic oscillator.  The harmonic oscillator in this case is not a mechanical harmonic oscillator but is defined in the space commonly known as the Fock space. 

Let us therefore consider the operators
\begin{eqnarray}\label{q621}
&{}& \hat{a}_{i} = \frac{1}{\sqrt{2}} \left(x_{i} + ip_{i} \right) =
\frac{1}{\sqrt{2}}\left(x_{i} +
\frac{\partial}{\partial x_{i} } \right), \nonumber\\[3ex]
&{}& \hat{a}^{\dagger}_{i} = \frac{1}{\sqrt{2}} \left( x_{i} - ip_{i} \right) =
\frac{1}{\sqrt{2}}\left(x_{i} - \frac{\partial}{\partial x_{i} } \right) .
\end{eqnarray}
Then
\begin{equation}
   \left[\hat{a}_{i}, \hat{a}^{\dag}_{j}\right] = \delta_{ij}  \qquad\mbox{and}\qquad
   \left[\hat{a}_{i}, \hat{a}_{j}\right] = 0 .
\end{equation}

For the case of one dimensional harmonic
oscillator the subscripts
$i$ and $j$ will be dropped.
The harmonic oscillator differential equation can be written as
\begin{equation} \label{co01}
	\frac{1}{2}\left\{-\left(\frac{\partial}{\partial x}\right)^2 +
	x^2\right\}\chi_{n}(x) = \left(n + \frac{1}{2}\right)\chi_{n}(x).
\end{equation}
As seen in Equation~(\ref{eq:346}), the solution of the harmonic oscillator wave
function in the $n^{\rm{th}}$ excited state is now written as
\begin{equation}\label{co02}
\chi_{n}(x) = \frac{1}{\sqrt{\pi 2^n n!}} H_{n}(x) e^{-x^2/2} \, .
\end{equation}
Here $H_{n}(x)$ is the Hermite polynomial.
If our
interest is in the quantum number $n$, it is more convenient to
write this wave function as $\ket{n}$.  Then
\begin{equation}\label{co03}
   \hat{a} \ket{n} = \sqrt{n} \ket{n-1} \,  \qquad\mbox{and} \qquad
     \hat{a}^{\dagger} \ket{n} =  \sqrt{n+1} \ket{n+1} \, .
\end{equation}
We call $\hat{a}$ and $\hat{a}^{\dagger}$ the step-down and step-up
operators, respectively, for the harmonic oscillator states.
In terms of these operators the number operator is
\begin{eqnarray} \label{eq622a}
\hat{N}=\hat{a}^{\dagger}\,\hat{a}  \qquad \mbox{with} \qquad \hat{N} \ket{n}=n\ket{n} \, .
\end{eqnarray} 
This algebra is the starting point for the Fock
space~\cite{fock34} 
where the number $n$ serves as the number of particles or photons in
a given state.  The operator $\hat{a}$ reduces the number of particles by
one, and $\hat{a}^{\dagger}$ adds one.  They are therefore known as the
annihilation and
creation operators, respectively.
These operators are 
the basic mathematical device for quantum optics.

In quantum optics, states 
consisting of one or two photons are important, as in the case of quantum 
electrodynamics.  However, equally important are both
coherent and  
incoherent mixtures of multi-photon states.

It is known that the uncertainty product is minimum for the 
ground-state harmonic oscillator or for the zero-photon state in Fock-space 
language.  The coherent state is a superposition 
of the multi-photon state  
which preserves the minimal uncertainty product, and which has a Poisson
distribution in photon numbers.  Since it has the minimum uncertainty, the 
wave function in Fock space takes a Gaussian form.
The Gaussian form can  
change its width without altering the uncertainty product~\cite{stoler1970}. 
This change in width may change the Poisson distribution to a different 
distribution.  The minimum-uncertainty states which do not have a Poisson 
distribution in photon number are called squeezed states.

The coherent state, represented here by $\ket{\alpha}$, is defined 
as~\cite{glauber63,sudar63,klauder85}
\begin{equation}\label{co547}
  \ket{\alpha} =  e^{(-|\alpha|^{2}/2)} 
  \sum_{n}\frac{\alpha^n}{\sqrt{n!}}\ket{n} \, .
\end{equation}
Here, the complex number $\alpha$ is written as
two real numbers $ Re(\alpha)= \alpha_a$ and $Im(\alpha)=\alpha_b $. 
This takes the form
\begin{equation}\label{co548}
\alpha = \alpha_a   + i\alpha_b \,  ,
\end{equation}
where $\alpha_{a}$ and $\alpha_{b}$ are related as
\begin{equation}\label{co611a}
\alpha_a = \sqrt{2} \, Re(\alpha) \quad\mbox{and}\quad  \alpha_b = \sqrt{2}\, Im(\alpha)\, .
\end{equation}
The coherent state is normalized
\begin{equation}
\braket{\alpha|\alpha} = 1 \, .
\end{equation}
The probability distribution for the photon number is
written as $P_{n}(n)=|\braket{n|\alpha}|^{2}$. Thus for the 
$n$-photon state the probability distribution is
\begin{equation}\label{co549} 
P_{n}(n) = \left(\alpha \alpha^{*}\right)^{n}\exp{\left(-\alpha
\alpha^{*} \right)}/n! \, .
\end{equation}
Therefore, the number of photons in the
coherent state has a
Poisson distribution.  
Ideal lasers can be formulated in terms of coherent states due to this 
very aspect of coherent states~\cite{arecchi65,knp91}.

Using the Wigner phase-space picture it is possible to study the coherent states of light. Here, the Wigner function is 
defined as~\cite{wig32a} 
\begin{equation}
 W(x,p) = \frac{1}{\pi} \int  e^{2ip\,x^{\prime}}
        \braket{x - x'|\alpha}^{*} \braket{x + x'|\alpha}~dx' \, ,
\end{equation}
in the two-dimensional space of
$x$ and $p$.  This integral can be evaluated with the resulting Wigner
function~\cite{bkn19iop}
\begin{equation}\label{co611}
W(\alpha; x,p) = \frac{1}{\pi}
      \exp\left\{-(x - \alpha_a)^{2} - (p - \alpha_b)^{2}\right\} \, .
\end{equation}

We note that this Wigner function produces the probability
distribution functions:
\begin{eqnarray}\label{eq:co559b}
\lvert  \braket{x|\alpha}\rvert^{2}  = \left(\frac {1}{\pi}\right)^{1/2}
\exp\left\{-\left(x - \sqrt{2}\alpha_{a} \right) ^{2} \right\} \, ,\\
\lvert \braket{p|\alpha}\rvert^{2}  = \left(\frac {1}{\pi}\right)^{1/2}
\exp\left\{-\left(p - \sqrt{2}\alpha_{b} \right) ^{2} \right\} \, ,
\end{eqnarray}
where $\alpha_{a}$ and $\alpha_{b}$ are as in Equation~(\ref{co548}).
Therefore, the probability distribution is seen to be a
Gaussian function for all possible
values of $\alpha$. 
The center of the distribution depends on the parameter $\alpha$. The real part of the center of distribution in the $\alpha$ variable, depends on the distribution in the $x$ variable and the imaginary part depends on the distribution in the $p$ variable.
If the $p$ variable is integrated out the probability distribution in the $x$ variable is:
\begin{equation}\label{co612}
\lvert \braket{x|\alpha}\rvert^{2}  = \int W(\alpha; x,p)~dp \, .
\end{equation}
Similarly, the probability distribution in the $p$ variable is given by
\begin{equation}\label{co612b}
\lvert \braket{p|\alpha}\rvert^{2}  = \int W(\alpha; x,p)~dx \, .
\end{equation}

The Wigner function of Equation~(\ref{co611}) leads to
\begin{eqnarray}\label{co613}
\lvert\braket{\beta |\alpha}\rvert^{2} & = &\int W(\beta;x,p)W(\alpha; x,p)~
dx~dp \nonumber \\
& = &\exp\left(-|\alpha - \beta |^{2} \right) \, .
\end{eqnarray}
This means that $\ket{\alpha}$ and $\ket{\beta}$ have overlaps and hence,
the coherent state is not a
complete orthonormal state, but over-complete.
If $\alpha = 0$, both $\alpha_a$ and $\alpha_b$ vanish, the Wigner function becomes
\begin{equation}\label{co614}
W(0;x,p) = \frac{1}{\pi}\exp\left\{-\left(x^{2} + p^{2} \right)\right\} \, ,
\end{equation}
which is concentrated within a circular region around the origin. This
is the Wigner function for the vacuum state. 

The exponentiation of operators and products of exponentiations are often involved in all branches of physics. To this end the Baker-Campbell-Hausdorff (BCH) relation is used~\cite{van2018}. 
The Wigner function of Equation~(\ref{co611}) can be obtained
from the vacuum-state Wigner function by making canonical 
transformations.
Explicitly, by making a translation along the $x$-axis by $\alpha_a$ and along the $p$-axis 
by $\alpha_b$ the Wigner function of Equation~(\ref{co611}) can be obtained from the vacuum state.
Hence, in the Wigner phase space, translations in the two
orthogonal directions, namely $x$ and $p$ are being used.  These generators in their differential forms 
can be written as
\begin{equation}  \label{co615}
P_{1}  = -i\frac{\partial}{\partial x} \,   \qquad\mbox{and}\qquad
P_{2} = -i\frac{\partial}{\partial p} \, .
\end{equation}
In this phase-space picture $P_{1}$  and $P_{2}$
commute with each other.
In addition, the rotation generator can be defined as
\begin{equation}\label{co616}
J_3 = -i~\left(x\frac{\partial}{\partial p} -
                    p \frac{\partial}{\partial x}\right) \, .
\end{equation}
In Section~\ref{sec:6} it is shown that $P_{1}$, $P_{2}$, and $J_3$ are the generators of the two-dimensional 
Euclidean group.  Rotations around the origin
form the one-parameter 
subgroup generated by $J_3$.  Translations generated by $P_{1}$ and
$P_{2}$ form a two-parameter Abelian invariant subgroup.  The translation operator
applicable to the Wigner function is
\begin{equation}\label{co618}
T(\alpha) = \exp\left(-\alpha_a\frac{\partial}{\partial x} -
 \alpha_b\frac{\partial}{\partial p} \right) \, ,
\end{equation}
which leads to
\begin{equation}
\label{co619}
T(\alpha) W(0;x,p) = W(\alpha; x,p) \, .
\end{equation}

Let us now write the transformation operators as
three-by-three matrices  applicable to the augmented space
$(1,~x,~p)$.  As seen in Equations~(\ref{eq:43}) and~(\ref{eq:45}), they can be written as
\begin{equation}\label{co623}
T(\alpha) = \begin{pmatrix}1 & 0 & 0 \cr \alpha_a & 1 & 0\cr \alpha_b & 0 & 1\end{pmatrix} \,  \quad \mbox{and} \quad
R(\theta) = \begin{pmatrix} 1 &0 &0 \cr 0 & \cos(\theta/2)  &  -\sin (\theta/2)  \cr 0 & \sin(\theta/2)  & \quad \cos (\theta/2) \end{pmatrix} .
\end{equation}
Then these matrices lead to
\begin{equation}\label{co624}
T(\beta) T(\alpha) = T(\alpha + \beta) 
\qquad \mbox{and} \qquad
R(\theta) T(\alpha) = T(\alpha') R (\theta).
\end{equation}
This demonstrates that the symmetry of the coherent
state is that of the two-dimensional Euclidean group in the Wigner
phase-space representation. 

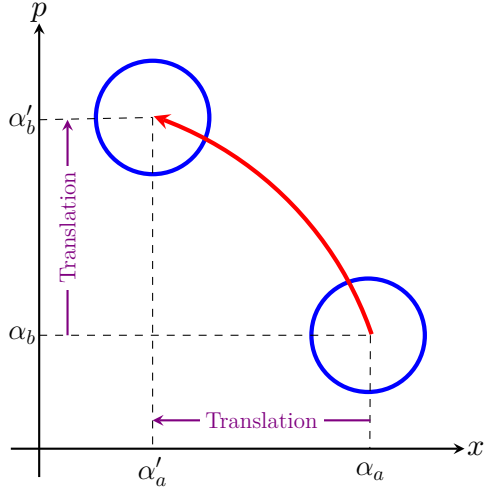
\begin{figure}[t!]
\centering
\begin{tikzpicture}[scale=0.75, transform shape]

 
\draw[black, thick, ->, >=stealth] (0.5,1) -- (8.5,1);
\draw[black, thick,->, >=stealth] (1,0.5) -- (1,8.5); 
        
\path (1, 8.7) node [black, scale=1.4] {$p$};
\path (8.7,1) node [black, scale=1.4] {$x$};
\draw[ultra thick, blue] (3,6.84) circle (1cm);
\draw[ultra thick, blue] (6.8,3) circle (1cm);

\draw[color=red, ultra thick, variable=\an, domain=19:71, ->, >=stealth]
        plot ({1+6.2*cos(\an)}, {1+6.2*sin(\an)});

\draw[dashed] (3,6.84) -- (3,1); 
\path (3, 0.6) node [black] {\Large${\alpha_{a}^\prime}$};
\draw[dashed] (6.84,3) -- (6.83,1);
\path (6.84,0.6) node [black] {\Large$\alpha_a$};
\draw[dashed] (1,6.8) -- (3,6.84);
\path (0.7,6.8) node [black] {\Large$\alpha_{b}^\prime$};
\draw[dashed] (1,3) -- (6.84,3);
\path (0.7,3) node [black] {\Large$\alpha_b$};

\draw[thick, ->, >=stealth, color=violet] (6.84, 1.5) -- node [fill=white,midway, scale=0.8] {\Large Translation} (3, 1.5);
\draw[thick, ->, >=stealth, color=violet] (1.5, 3) -- node [fill=white,midway, scale=0.8, rotate=90] {\Large Translation} (1.5, 6.8);

\end{tikzpicture}

\caption{For a Euclidean transformation in phase space, a rotation around the origin results in two successive translations~\cite{bkn19iop}.}\label{cof012}
\end{figure}


The Wigner function of Equation~(\ref{co611}) is localized in the
circle:
\begin{equation}\label{co617}
(x - \alpha_a)^{2}  + (p - \alpha_b)^{2}  = 1 \, .
\end{equation}
Therefore, all the instruments developed in Section~\ref{sec:4} for
canonical transformations of a circle
are now applicable to the Wigner function for the
coherent state.

Multiplying $\alpha$ by $e^{i\theta}$, the circle of
Equation~(\ref{co617}) is rotated around the origin, with the
resulting equation
\begin{equation}\label{co620}
(x - \alpha_{a}^{\prime})^{2}  + (p - \alpha_{b}^{\prime})^{2} = 1 \, ,
\end{equation}
where
\begin{equation} \label{eq:matphi}
\begin{pmatrix}\alpha_{a}^{\prime} \cr\alpha_{b}^{\prime} \end{pmatrix} =
\begin{pmatrix}\cos\theta & -\sin\theta \cr \sin\theta & \;\;\; \cos\theta \end{pmatrix}
\begin{pmatrix}\alpha_a \cr \alpha_b \end{pmatrix} \, .
\end{equation}

As shown in Figure~\ref{cof012}, the rotation $R(\theta)$ results in
only a translation of the circle centered at the origin to $(\alpha_{a}^{\prime},\alpha_{b}^{\prime})$. It should be noted that the translated vacuum represented by a circle at the origin is a coherent state.

\subsection{Single-Mode Squeezed States of light}\label{sec:63}

The squeeze and rotation generators used here are those given in 
Equation~(\ref{eq:416a}). These generators are provided in Table~\ref{tabl1} 
and they satisfy 
the commutation relations that form the Lie algebra for the $Sp(2)$ group. 
When applied to the Wigner function the rotation generator has the form:
\begin{equation} 
J_2 = -\frac{ i}{2} \left(  x \frac{ \partial }{ \partial p } -  p\frac{
\partial }{ \partial x} \right) \, . \label{eq:616} 
\end{equation} 
The squeeze generators take the form 
\begin{equation} 
K_{1} = - \frac{i}{ 2} \left( x\frac{ \partial }{
\partial x} - p \frac{ \partial }{ \partial p} \right) ,\qquad
K_{3} =  \frac{i}{2} \left( x\frac{ \partial }{
\partial p} + p\frac{ \partial }{ \partial x } \right) \, . \label{eq:628} 
\end{equation} 
They satisfy the same commutation relations in phase space as given in 
Equation~(\ref{eq:416a}).

For the squeezed state the Wigner function takes the following form: \begin{equation} 
W(\alpha, \lambda;x,p) = \frac{ 1}{\pi} \exp \left\{ - \left[
e^{-\lambda} (x - e^{\lambda /2} \alpha_a)^{2} + e^{\lambda} (p - e^{-\lambda
/2} \alpha_b)^{2} \right] \right\} \, .\label{eq:629} 
\end{equation} 
We can obtain this by applying the squeeze operator
\begin{equation} 
S(\lambda) = \exp \left\{ -\frac{\lambda}{2} \left(x\frac{ \partial
}{\partial x} - p\frac{ \partial }{ \partial p} \right) \right\}
,\label{eq:630} 
\end{equation} 
to the Wigner function for a coherent state given in Equation~(\ref{co611}).
Unlike in the
case of coherent states, the distribution for a
squeezed state is elliptic.   
The area in phase space of this ellipse is the same as that of the circle.  The
squeezed state is also a minimum uncertainty state.

Suppose the squeeze is made along the direction that makes an angle 
$(\theta /2)$ with the $x$-axis, then the squeeze operator is 
\begin{eqnarray} 
S(\theta, \lambda) & = &  R(\theta )S(0,\lambda )R(-\theta ) \nonumber \\[2mm]
\nonumber  \\
 & = &  \exp \left\{  -\frac{ \lambda }{2} \left[ \sin \theta
\left( 
x \frac{ \partial }{ \partial x } - p\frac{ \partial }{ \partial p}
\right)  - \cos \theta  \left( 
x\frac{ \partial }{ \partial p} + p\frac{\partial }{ \partial x}
\right) \right]\right\} \, . 
\label{eq:632} 
\end{eqnarray} 
In view of Equation~(\ref{eq:628}), 
we have:
\begin{equation} 
R(\theta ) S(0,\lambda)  R(-\theta ) = \exp \left \{- i\lambda \left[ \sin \theta \,
  K_{1} + \cos \theta \, K_{3} \right] \right \}\, . 
\label{eq:631} 
\end{equation} 
Therefore, two parameters are needed, namely $\lambda$ and
$\theta$.  These  parameters can be given by one complex variable
$\zeta$, where $\zeta$ is defined as $\lambda e^{i\theta}$ and $0\leq\theta < 2\pi$ describes the squeezing direction.  Here, $\lambda$ describes the strength of the squeezing, with $e^{\lambda} > 0 \quad \mbox{and} \quad e^{-\lambda} > 0 \quad \mbox{for all real } \lambda$. The single squeezed
state $\ket{\zeta}$ is obtained from
\begin{equation}\label{eq:631a}
\ket{\zeta} = S(\zeta)\ket{0} \, ,\quad \ket{\zeta\, ,\alpha} = S(\zeta)\ket{\alpha} \, .
\end{equation}

Squeezes alone do not form a group, rotations must be added to form a group.
Two such groups have already been mentioned, namely, the Lorentz
group $SO(2\, ,1)$ and the symplectic group $Sp(2)$. 
The matrix form of the squeeze operator, where we use Equation~(\ref{eq:45}) for $R(\theta)$, is  identical to that of Equation~(\ref{eq:139}). 
This is applicable to the column vector of $(1, x, p)$.

As was seen in Section~\ref{sec:6}, the group of linear canonical
transformations
in two-dimensional phase space consists of translations, rotations, and
squeezes.  Translations and rotations can form a group, for example the
two-dimensional Euclidean
group $E(2)$. Equally, rotations and squeezes can form a group, for example, 
$SO(2\, ,1)$ and $Sp(2)$.  Translations form their own
subgroup.  It was shown in Section~\ref{sec:6} that translations form an 
invariant subgroup.  Now, we wish to examine squeezes and translations.
A simple matrix algebra leads to 
\begin{equation} 
S(\zeta ) T(\alpha ) \left[  S(\zeta) \right] ^{-1}   = T(\alpha^{\prime\prime}) \, ,
\label{eq:634} 
\end{equation} 
with
\begin{eqnarray}  
\alpha_a^{\prime\prime} & = &  \left(  \cosh \frac{ \lambda}{2} + (\cos \theta )\sinh\frac{
\lambda}{2} \right) \alpha_a +  \left(  ( \sin \theta )\sinh\frac{ \lambda}{2}
\right) \alpha_b , \nonumber \\[2mm]
\alpha_b^{\prime\prime} & = & \left(  (\sin \theta )\sinh\frac{\lambda}{2}\right) \alpha_a + \left(
\cosh\frac{\lambda}{2}  - (\cos \theta )\sinh\frac{ \lambda}{2} \right) \alpha_b \, .
\end{eqnarray}
Hence, the translation does not commute with squeezes.  However, once
commuted, it still remains an element of the translation
subgroup even
though its parameters are changed.  This is a manifestation of the fact
that the translation subgroup is an
invariant subgroup. 
Equation~(\ref{eq:634}) can be written as 
\begin{equation}
T(\alpha^{\prime\prime}) S(\zeta ) = S(\zeta ) T(\alpha ) .\label{eq:635} 
\end{equation} 

The product of squeezes, rotations, and translations constitute
the most general form of canonical
transformations in quantum mechanics. Hence,
from Equation~(\ref{eq:632}) and  
Equation~(\ref{eq:635}), 
the most general form is written as a translation followed by
a squeeze and a rotation.  If written as $TSR$ and operated on the
vacuum state, $R$ does not have any effect.  Thus, a translation of a squeezed vacuum $TS$ is the most general form
applicable to the vacuum.

\subsection{Squeezed Vacuum}\label{sec:64}

\begin{figure}[!t] 
\begin{center}
\begin{tikzpicture}[scale=0.75, transform shape]
\path[inner sep=0] node at (2,0) {\includegraphics[width=0.5\textwidth]{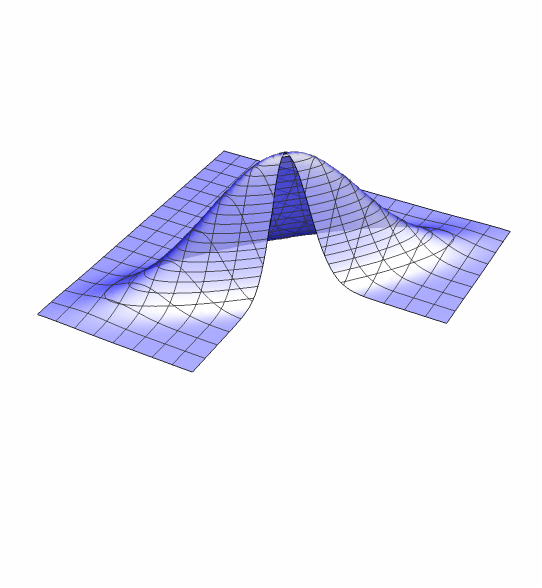}};
\path[inner sep=0] node at (12,0) {\includegraphics[width=0.5\textwidth]{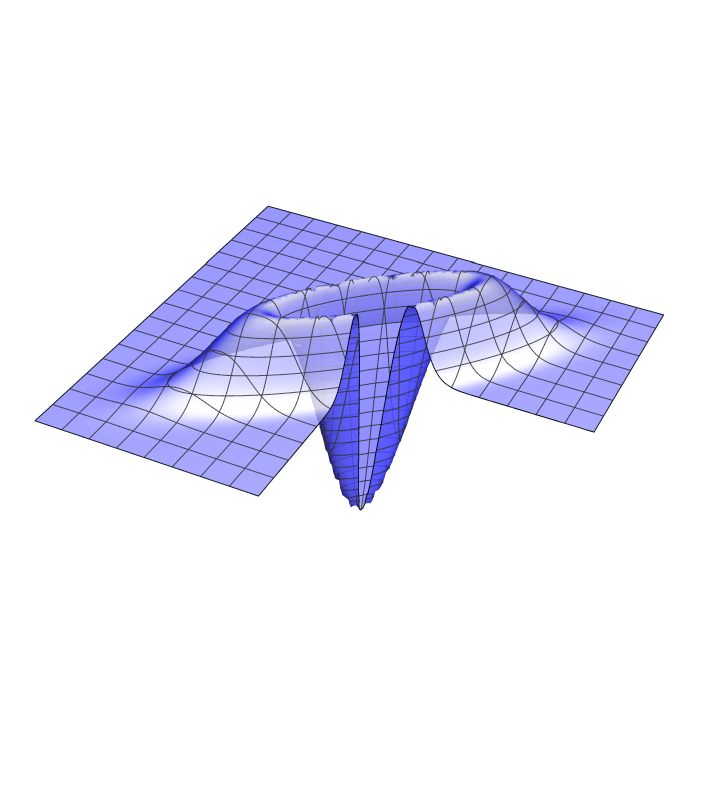}};
\path (3.6,-1.4) node [black] {\Large$W^{(\mathrm{sq})}_{0}(x,p)$ };
\path (14.4,-1.4) node [black] {\Large$W^{(\mathrm{sq})}_{1}(x,p)$ };


\end{tikzpicture}
\end{center}
\vspace{-2.4cm}
\caption{The Wigner function for the squeezed vacuum and the first excited squeeze. Here,\\
$\hspace*{4mm}W^{(\mathrm{sq})}_{0}(x,p)=
\frac{1}{\pi}
\exp \left\{-(1/2) \left[ e^{-\lambda}(p+x)^{2}+e^{\lambda}(p-x)^{2} \right] \right\}\,$ and \\
$\hspace*{4mm}W^{(\mathrm{sq})}_{1}(x,p)=\frac{1}{\pi}\exp \left\{-(1/2) \left[ e^{-\lambda}(p+x)^{2}+e^{\lambda}(p-x)^{2} \right] \right\}
\left[e^{-\lambda}(p+x)^{2}+e^{\lambda}(p-x)^{2}-1  \right]$, \\where we have chosen $\sin
\frac{\theta}{2}=\cos
\frac{\theta}{2} $ and $e^{\lambda}=4$ in the plot.
When compared with the ground and the first excited state of the Wigner 
functions in Figure~\ref{fig:31}, it is seen now that the axial symmetry is lost, and the effect of the squeeze is apparent through the elliptical form of the distribution function.}  \label{fig:sqw0w1}
\end{figure}

The study of
squeezed states as concluded in Section~\ref{sec:63} therefore starts from the squeezed vacuum. The squeezed vacuum will be studied in 
detail now.  The form of the Wigner function
for the squeezed vacuum is
\begin{equation} 
W(0, \lambda;x,p) =  \frac{1}{\pi} \exp \left\{  -e^{-\lambda}
\left(  x 
\cos \frac{\theta}{2} + p \sin \frac{\theta }{2} \right) ^{2}  - e^{\lambda} \left( - x \sin
\frac{\theta}{2} + p \cos 
\frac{ \theta }{ 2} \right) ^{2} \right\} \, . \label{eq:636} 
\end{equation} 
It is a function of $\lambda$ and $\theta$. The squeezed vacuum and the first excited squeezed state is depicted in Figure~\ref{fig:sqw0w1}.

Suppose now that we have a function $F(x,p)$ defined in the Wigner phase space. The expectation value of
$F(x,p)$ is  
\begin{equation} 
\hspace{5mm} \braket{F}_{0} =\frac{1}{\pi} \int F(x,p) \exp \left\{ 
-e^{-\lambda} \left(  x \cos \frac{\theta}{2} + p \sin \frac{\theta}{2}
\right) ^{2}
 -e^{\lambda} \left( - x \sin \frac{\theta}{2} + p \cos
\frac{ \theta}{2} \right) ^{2} \right\}~dx~dp \, . \label{eq:637} 
\end{equation}
The subscript zero is used to indicate that the expectation value is taken 
for the squeezed vacuum.  Because the volume element $dxdp$ is invariant under 
canonical transformations, the integral is also invariant under canonical  
transformations. If the coordinate is rotated to align 
the major axis of the vacuum state with the $x$ or $p$ axis, then
\begin{equation} 
\braket{F}_{0}  = \frac{ 1}{\pi} \int  F(x',p') \exp \left\{   - \left(x
^{2} + p^{2} \right) \right\}  ~dx ~dp \, , \label{eq:638} 
\end{equation}
where          
\begin{equation} 
x^{\prime}  = \left(  e^{\lambda /2} \cos \frac{\theta}{2}\right)  x -  \left(
e^{-\lambda /2}     \sin \frac{\theta}{2} \right) p \, , \qquad
p'  =  \left(  e ^{-\lambda /2}  \sin \frac{ \theta}{2} \right) x +  \left(
e^{\lambda /2}   \cos \frac{ \theta}{2} \right)  p \, .
\end{equation} 
Therefore,  $\braket{x}_{0}$  and $\braket{p}_{0}$  are zero, but
\begin{equation} 
\braket{x^{2}}  =  \frac{ \cosh \lambda + (\sinh\lambda)\cos\theta}{2},
\qquad 
\braket{p^{2}}  =  \frac{ \cosh \lambda -
(\sinh\lambda)\cos\theta}{2} \, . \label{eq:639} 
\end{equation} 

As a consequence
\begin{equation} 
\braket{x^{2}}\braket{p^{2}} = \frac{1 + (\sin \theta)^{2} ( \sinh \lambda) ^{2}}{4} \,  . \label{eq:640} 
\end{equation}  
This result is the same as obtained in the Schrödinger picture,  
frequently mentioned in the literature~\cite{caves85,henry88,fisher84}.
The above quantity can not be greater 
than 1/4 in order for the squeezed state 
to be a minimal uncertainty state. It is possible to resolve this problem.
\begin{figure}[htb]
\centering
\begin{tikzpicture}[scale=0.90, transform shape]

\draw[black,  thick, ->, >=stealth] (1,3) -- (5,3);
\draw[black,  thick,->, >=stealth] (3, -0.2) -- (3, 6.2); 

\draw[violet, thick, <->, >=stealth , rotate around={65.0:(3,3)}] (-0.4, 3) -- (6.4, 3);
\draw[violet, thick, ->, >=stealth, rotate around={155.0:(3,3)}] (1, 3) -- (2.6, 3);
\draw[violet,  thick, <-, >=stealth , rotate around={335.0:(3,3)}] (2.6, 3) -- (1, 3) ;


\draw[ultra thick, blue] (3,3) circle (1cm);

\draw[ultra thick,red, rotate around={65.0:(3,3)}] (3,3) ellipse (3 cm and 0.333 cm);

\path (5.2,3) node [black] {$x$};
\path (3,6.4) node [black] {$p$};  

\path (5.7,1.8) node [violet] {Squeezing};

\end{tikzpicture}
\caption{Pictured is the squeezed vacuum and vacuum states.
Here, the circle  
corresponds to the vacuum state, which is a coherent state, and the tilted ellipse is for the 
squeezed vacuum.  The squeezed vacuum is not a zero-photon state, but 
is a minimum uncertainty state in the sense that the area of the 
ellipse is the same as that of the circle.  This interpretation 
is similar to the case of wave-packet spreads discussed in
Section~\ref{sec:3}.}
\label{fig:62}
\end{figure}
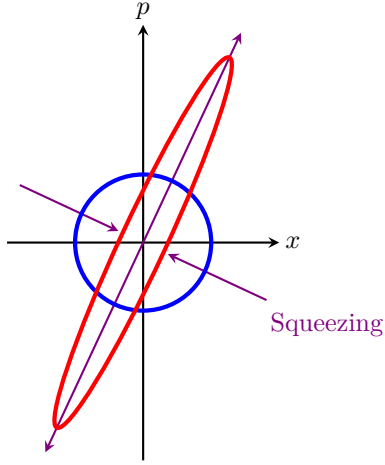
For this purpose, we refer to Figure~\ref{fig:62}. Canonical transformations in quantum mechanics are allowed in the phase-space picture.
The rotation of $W(0,\lambda ;x,p)$ to 
\begin{equation} 
W(0,\lambda ;x,p) =  \frac{ 1}{\pi}\exp \left\{ - \left(  e^{-\lambda}
x^{2}  + e^{\lambda}  p^{2} \right) \right\} \label{eq:641} 
\end{equation} 
is indeed a canonical transformation in
quantum mechanics~\cite{hkn89}.  This expression gives the minimal uncertainty product.  
It is now possible to quantify
in phase space the uncertainty in terms of the area where the Wigner
function is concentrated. This is very similar to the
wave packet spread where the amount of
uncertainty is constant in time,  
while, in the Schrödinger picture, the uncertainty product
increases as time
progresses or regresses. The wave packet
spread was discussed in detail in
Section~\ref{sec:3}. 

The photon number and
the (photon number)$^{2}$ operators are, of all the operators in quantum mechanics, among the two most important operators in
quantum optics.  The number operator in the Schrödinger picture has
the form
\begin{equation} 
\hat{N} = \frac{1}{2} \left\{  x^{2}  -  \left(  \frac{ \partial }{
\partial x} \right) ^{2}   - 1 \right\}  \, .\label{eq:642} 
\end{equation} 
Hence in phase space, the number operator is 
\begin{equation} 
\hat{N} = \frac{1}{2} \left(  x ^{2} + p^{2}  - 1 \right)  \, . \label{eq:644} 
\end{equation}
Therefore, the expectation value, in the phase-space picture, of this operator is 
\begin{equation} 
\braket{\hat{N}} = \frac{1}{2} \int \left(   x^{2}  + p^{2}  - 1 \right)  W(x,p)~dx~dp \, . \label{eq:643} 
\end{equation} 
The multiplication of two $\hat{N}$ number operators is not straightforward in the quantum phase-space, thus for $\hat{N}^{2}$, we resort to the Moyal product~\cite{moyal_1949,curtright_2014,marino_2021}, which in our case is $\hat{N} \star \hat{N}$. Then we obtain 
\begin{equation}
\hat{N}^{2}  = \frac{1}{4} \left\{ \left(  x^{2}  + p^{2}  - 1 \right) ^{2}-1
\right\} \, ,\label{eq:646} 
\end{equation} 
to evaluate the expectation value
\begin{equation} 
\braket{\hat{N}^{2}}= \frac{ 1}{4} \int \left\{ \left(   x ^{2} + p^{2}  - 1 \right) ^{2} -1 \right\} W(x,p)~dx~dp  \, .\label{eq:645} 
\end{equation}
Consider next $\braket{\hat{N}}_{0}$ which from the expressions
given in Equations~(\ref{eq:636}) and~(\ref{eq:644}) can be written as
\begin{equation} 
\hspace{2mm} \braket{\hat{N}}_{0} =   \frac{1}{2\pi} \int \left(  x^{2}  + p^{2}  - 1 \right)   \exp \left\{ -e^{-\lambda} \left(  x \cos  \frac{ \theta}{2} +
p \sin \frac{ \theta}{2} \right) ^{2}
  - e^{\lambda}  \left( - x \sin
\frac{ \theta}{2}+ p \cos 
\frac{\theta}{2} \right) ^{2} \right\}~dx~dp \, .  \label{eq:647} 
\end{equation} 
As $\left(x ^{2} + p ^{2} - 1 \right) $ is invariant under
rotations, the integral in Equation~(\ref{eq:647}) can be  
reduced to 
\begin{equation} 
\braket{\hat{N}}_{0}  = \frac{1}{2\pi} \int \left(   x^{2}  + p^{2}  - 1 \right)
\exp \left\{-  \left(  e^{-\lambda}   x^{2}  + e^{\lambda} p ^{2}
\right) \right\}~dx~dp \, . \label{eq:648} 
\end{equation} 
Equation~(\ref{eq:648}) can be written now as the vacuum expectation value of
$\frac{1}{2} \left(   e^{\lambda }x^{2}  + e^{-\lambda}  p^{2}  - 1
\right) $:
\begin{equation} 
\braket{\hat{N}}_{0} = 
\frac{1}{2\pi}\int
\left(   e^{\lambda}  x^{2}  + 
e^{-\lambda}   p^{2}   - 1 \right)  \exp \left\{ - \left(  x^{2}  +
p^{2} \right) \right\}~dx~dp \, . \label{eq:649} 
\end{equation} 
The result is 
\begin{equation} 
\braket{\hat{N}}_{0}  = \frac{1}{2} \left(  \cosh \lambda -1 \right) \, .\label{eq:650} 
\end{equation} 
This Equation~(\ref{eq:650}) vanishes when $\lambda = 0$, showing that, in an unsqueezed vacuum, the number of photons is zero.  

Now $\braket{\hat{N}^{2}}_{0}$ is calculated, as $\hat{N}^{2}$ is invariant under rotations, 
\begin{equation} 
\braket{\hat{N}^{2}}_{0} = 
\frac{1}{4\pi} \int \left\{  \left(   e^{\lambda} x^{2}
+ e^{-\lambda }  p^{2}  - 1 \right) ^{2} - 1 \right\} \exp
 \left\{
- \left(  x^{2}  + p^{2} \right)  \right\}~dx~dp \, .
\label{eq:651} 
\end{equation} 
The evaluation of this integral leads to 
\begin{equation} 
\braket{\hat{N}^{2}}_{0}  =  \frac{3}{8}\cosh(2 \lambda)   - \frac{1}{2}\cosh
\lambda +\frac{1}{8}  \, . \label{eq:652} 
\end{equation}
Equation~(\ref{eq:652}) vanishes when $\lambda = 0$ since $\braket{\hat{N}^{2}}_{0}$ is seen to be independent of the direction of the squeeze.  From this we can calculate 
\begin{equation} 
\braket{(\Delta \hat{N})^{2}}_{0} = \braket{\hat{N}^{2}}_{0} - \braket{\hat{N}}^{2}_{0}\, . \label{eq:delta0}
\end{equation}
From Equations~(\ref{eq:650}) and~(\ref{eq:652}), we have
\begin{equation} 
\braket{(\Delta \hat{N})^{2}}_{0} = \frac{1}{2}(\cosh^2\lambda -1) = \frac{ 1}{2}\sinh^{2}\lambda  \, . \label{eq:653}
\end{equation} 

Within the framework of the
Schrödinger picture, it is possible to derive this same result.
The advantage of using the
phase-space picture is that it allows the use of symmetry properties in 
phase space.  For example, let us go back to Figure~\ref{fig:62}.  In integrating
over the entire phase space, we can use the coordinate system in which the major or minor axis of the ellipse coincides with the coordinate axis.  The 
integration is also invariant under translations in phase space.  

\section{Symmetries of Two-Mode States}\label{sec:68}

The two-mode coherent photon state
has occupied a central role in optical sciences since 1976~\cite{yuen76}.
States having two different photons have a pivotal role in quantum physics.
They are variously called two-photon coherent states, squeezed states, or two-photon
entangled states. The two-photon state can be denoted as
\begin{equation}\label{sq101}
\ket{n_{1}}\ket{n_{2}}=\ket{n_{1},n_{2}} \,  \, .
\end{equation}
The most general form for this two-mode state often takes the form
\begin{equation}\label{sq102}
\sum^{}_{n_{1},{n_{2}}}  A_{n_{1},n_{{2}}}  \ket{n_{1},n_{2}} \, .
\end{equation}
When $n_{1}=n_{2}=n$, it is conventionally expressed as:
\begin{equation}
    \ket{\chi}=\exp\{\chi(\hat{a}_{1}^\dagger \hat{a}_{2}^\dagger-\hat{a}_{1}\hat{a}_{2})\}\ket{0,0}.
\end{equation}
Here $\exp\{\chi(\hat{a}_{1}^\dagger \hat{a}_{2}^\dagger-\hat{a}_{1}\hat{a}_{2})\}$ is a unitary operation on the state,
whose generator is 
\begin{equation}
    Q_{3}=\frac{i}{2}(\hat{a}_{1}^\dagger \hat{a}_{2}^\dagger-\hat{a}_{1}\hat{a}_{2})
\end{equation}
in terms of the operators in Fock space. We shall not go any further about the treatment of the symmetries of two-mode or squeezed states in Fock space, which can readily be found in~\cite{bkn19sym} since our main focus is on the Wigner functions and on their symmetry properties.  Nevertheless, for completeness, we give
in Appendix B the equivalences of these generators in both spaces, i.e., in phase space and in Fock space.

It was shown in Section~\ref{sec:6}
that the algebraic property of the
group of homogeneous linear canonical transformations in quantum mechanics
is the same as that
of the Lorentz group $SO(2\, , 1)$ and the symplectic group $Sp(2)$. 

We shall see that the basic symmetry of 
two-mode squeezed states is that of
$SO(3\, ,2)$. In the laboratory, most squeezed states produced are
two-mode states. Nevertheless, for practical applications, specifically when applied in laboratories and technological environments, the basic symmetry of interest
is still $SU(1\, ,1)$~\cite{yurke86,gilles_nonclassical_1992,gerry_two-mode_2000}, which is isomorphic 
to $SO(2\, ,1)$~\cite{chai_two-mode_1992,campo_inflationary_2005,ren_entanglement_2019,ren_nonclassicality_2019}. 

\subsection{Wigner functions and the Symmetries of Two-Mode
States}\label{sec:91}

To define the symmetry of two different modes, we need the
four-dimensional Wigner function given, in a slightly unconventional order, by
\begin{equation}\label{eq:688a}
 W(x_{1}, x_{2}, p_{1}, p_{2} ) \, .
\end{equation}
Our interest focuses on the Wigner function for
the two-mode vacuum state:   
\begin{equation} 
W_0(x_{1}, x_{2}, p_{1}, p_{2} ) = \frac{ 1}{\pi^{2}} \,
\exp \left\{ -\left( x_{1}^{2} + x_{2}^{2}  + p_{1}^{2}  + p^{2}_{2}
\right) \right\} \, . \label{eq:688} 
\end{equation} 
Here, we note that the
four-dimensional phase
space can therefore be divided into two two-dimensional spaces in many different ways as we shall see below. 

The Wigner phase space representation of
quantum mechanics~\cite{wig32a,knp91} allows a more thorough
study of the symmetry problems of the two-mode state.
The Wigner phase-space distribution function in
this two-mode state is defined as 
\begin{equation} \label{sq501}
W\left(x_{1}, x_{2}, p_{1}, p_{2} \right)
 = \frac{1}{\pi^{2}}\int e^{2i\left(p_{1}y_{1} + p_{2}y_{2}\right)}\, \rho\left(x_{1} - y_{1}, x_{2} - y_{2};
 x_{1} + y_{1}, x_{2} + y_{2}\right) ~dy_{1}~ dy_{2}\, ,
\end{equation}
where
the density matrix is defined as~\cite{neum2018}:
\begin{equation}\label{sq503}
\rho(x_{1}, x_{2}; x'_{1}, x'_{2}) =
\psi\left(x_{1}, x_{2}\right) \psi^{*} \left(x'_{1}, x'_{2}\right) \, .
\end{equation}
If the $x_{2}$ variable is not observed, this density matrix becomes
\begin{equation} \label{sq504}
\rho(x_{1}, x'_{1}) = \int \rho(x_{1}, x_{2}; x'_{1}, x_{2})~dx_{2} \, ,
\end{equation}
and the trace of this density matrix is one:
\begin{equation}
\int \rho(x, x) dx  = 1 \ .
\end{equation}

This two-mode state has a vacuum state that corresponds to
the two-harmonic oscillator system. The ground state of the
two-harmonic oscillator system has the wave function:
\begin{equation}\label{sq505}
\psi\left(x_{1}, x_{2}\right)= \frac{1}{\sqrt{\pi}}
     \exp \left \{-\frac{1}{2}\left({x_{1}^2 + x_{2}^2}\right) \right \} \, .
\end{equation}
Thus, the density matrix is
\begin{equation} \label{sq507}
\rho(x_{1}, x_{2}; x'_{1}, x'_{2})
 = \frac{1}{\pi} \exp \left \{-\frac{1}{2}\left(x_{1}^{2} + x_{2}^{2} +
      (x'_{1})^{2} + (x'_{2})^{2}\right) \right \} \, .
\end{equation}
The Wigner function corresponding to the ground state is presented in Equation~(\ref{eq:688}), where it is defined in the four-dimensional phase space.
A set of rotation operators can be defined as
\begin{eqnarray}\label{eq:689}
J_{1} & = &+   \frac{ i}{2} \left\{ \left( x_{1}
\frac{\partial }{\partial p_{2}} - p_{2} \frac{ \partial }{\partial 
x_{1}} \right) +  \left(  x_{2} \frac{\partial }{\partial p_{1}} - p
_{1}\frac{ \partial }{\partial x_{2}} \right) \right\} \, , 
\nonumber \\[2mm]
J_{2} & = & -   \frac{i}{2}  \left\{ \left(  x_{1}
\frac{\partial }{\partial x_{2}} - x_{2} \frac{\partial }{\partial
x_{1}}  \right) + \left(   p_{1} \frac{\partial }{\partial p_{2}} -
p_{2}\frac{ \partial }{\partial p_{1}} \right) \right\} \, ,
\nonumber \\[2mm]
J_{3}  & = &  +    \frac{i}{2}  \left\{ \left(  x_{1}
\frac{\partial }{\partial p_{1}} - p_{1} \frac{\partial }{\partial
x_{1}} \right)  - \left(   x_{2} \frac{\partial }{\partial p_{2}} -
p_{2} \frac{\partial }{ \partial x_{2}} \right ) \right \} \, , 
\nonumber \\[2mm]
J_{0} & = & +   \frac{i}{2}  \left\{ \left(   x_{1}
\frac{\partial }{\partial p_{1}} - p_{1} \frac{\partial }{\partial
x_{1}} \right)  + \left( x_{2} \frac{\partial }{\partial p_{2}} -
p_{2} \frac{ \partial }{\partial x_{2}} \right) \right\} \, . 
\end{eqnarray} 
$J_1$ generates separated rotations in the spaces of $(x_{1}p_{2})$ and $(x_{2} p_{1} )$. The rotations are in the same direction. 
$J_2$ generates rotations in the spaces of $(x_{1}x_{2})$ and $(p_{1} p_{2} )$ in the same direction.
$J_{0}$ and $J_{3}$ generate rotations the $(x_{1} p_{1} )$ and $(x_{2} p_{2} )$
spaces.  The rotations are in
the same direction for $J_{0}$ and in the opposite directions for $J_{3}$.

It is possible to
rewrite the boost operators $K_{1}$, $K_{2}$, and $K_{3}$ as 
\begin{eqnarray}\label{eq:692}
K_{1}  & = & -\frac{ i}{2}\left\{ \left( x_{1} \frac{
\partial }{\partial p_{1}} + p_{1} \frac{\partial }{\partial x_{1} }
\right)   - \left(   x _{2}\frac{\partial }{\partial p_{2}} + p_{2}
\frac{\partial }{\partial x_{2} }\right) \right\} \, , \nonumber \\[2mm]
K_{2} & = &  -   \frac{i}{2}  \left\{ \left(   x_{1}
\frac{\partial }{\partial x_{1}} - p _{2}\frac{\partial }{\partial
p_{2}} \right)  + \left(   x_{2} \frac{\partial }{\partial x_{2}} -
p_{1} \frac{\partial }{\partial p_{1}} \right) \right\} \, ,\nonumber \\[2mm]
K_{3} & = &  +  \frac{i}{2}  \left\{ \left(   x_{1}
\frac{\partial }{\partial p_{2} }+ p_{2} \frac{\partial }{\partial
x_{1}} \right)  +  \left(  x_{2} \frac{\partial }{\partial p_{1}} +
p_{1} \frac{ \partial }{\partial x_{2}} \right) \right\} \, . 
\end{eqnarray}
Now $J_{1}$, $K_{2}$, and $K_{3}$ satisfy the set of commutation
relations for $(2 + 1)$-like transformations in the $(x_{1} p_{2} )$ and
$(x_{2} p_{1} )$ 
spaces separately.  Likewise, the subgroups generated by
$J_{2}, K_{3}, K_{1}$ and by
$J_{3}, K_{1}, K_{2}$ perform $(2 + 1)$-like transformations in two separate
two-dimensional spaces. 

To obtain the complete symmetry for the two-mode squeezed states
consisting of the Lorentz
group $SO(3 \, ,2)$, also known as the (3+2) de Sitter group, it is necessary to include one more set of three boost generators, $Q_1$, $Q_2$, and $Q_3$. 
These generators have the form:
\begin{eqnarray}\label{eq:695}
Q_{1} & = & +    \frac{i}{2}  \left\{ \left(  x_{1}
\frac{\partial }{\partial x_{1}} - p _{1}\frac{\partial }{\partial
p_{1}}\right) - \left(   x _{2}\frac{\partial }{\partial x_{2}} -
p_{2} \frac{\partial }{\partial p_{2}}\right) \right\} \, , \nonumber\\[2mm]
Q_{2}  & = & -    \frac{ i}{2} \left\{ \left(   x_{1}
\frac{\partial }{ \partial p_{1} } + p_{1} \frac{\partial }{\partial
x_{1}} \right)  + \left(  x_{2} \frac{ \partial }{\partial p_{2}} +
p_{2} \frac{\partial }{ \partial x_{2}} \right) \right\} \, , \nonumber\\[2mm] 
Q_{3}  & = & -    \frac{i}{2}  \left\{ \left( x_{2} \frac{\partial }{\partial x_{1}} + x_{1} \frac{\partial }{ \partial x_{2}}\right)   -
\left(   p_{2} \frac{ \partial }{ \partial p_{1}} + p_{1}
\frac{\partial }{ \partial p_{2}} \right) \right\} \, .
\end{eqnarray}
Their commutation
relations with the rotation generators $J_i$ are given by:
\begin{equation} 
[Q_{i} , Q_{j} ] = -i\varepsilon _{ijk} J_{k}  \, , \qquad  [J_{i} , Q_{j} ] = i\varepsilon _{ijk}    Q_{k}  \, , \label{eq:693} 
\end{equation} 
for $J_{1}, J_{2}$, and $J_{3}$.  Transformations generated by these operators are also
like those of the $SO(3\, ,1)$ Lorentz group. See Appendix A for the formal definition of the the $SO(3\, ,1)$ Lorentz group. These
transformations are not separable into two two-dimensional phase spaces.
However, it is possible to transform
each $Q_{i}$ into
$K_{i}$ by a rotation generated by $J_{0}$, while the $J_{i}$ remain
invariant under the same rotation.

It is now possible to consider the $SO(2\, ,1)$-like subgroup generated by
$J_{0}, K_{1}$,  
and $Q_{1}$ where this subgroup also performs transformations in two
separate two-dimensional phase spaces.  As for
$SO(2\, ,1)$ groups $J_{0}, K_{2}, Q_{2}$, and the $J_{0}, K_{3}, Q_{3}$ subgroups, $J_{0}$ is separable
in $(x_{1} p_{1} )(x_{2} p_{2} )$, while $K_{2}$ 
and $Q_{2}$ as well as and $K_{3}$ 
and $Q_{3}$ are separable in $(x_{1} p_{2} )(x_{2} p_{1} )$. There are altogether
nine $SO(2\, ,1)$-like subgroups
in the $SO(3\, ,2)$-like symmetry group of two-mode
squeezed states.  They are either separable or
can be transformed into separable representations.  

The ten generators defined in this section for the group $SO(3\, ,2)$ also serve as the generators for the group $Sp(4)$. In his 1963 paper~\cite{dir63}, Dirac defined the ten generators of $Sp(4)$, $J_i$, $K_i$, and $Q_i$ where $i$ equals $1,~2,~3$ with the addition of $J_0$, in terms of the annihilation and creation operators $\hat{a}$ and $\hat{a}^{\dagger}$. For the relation of the phase space generators defined in Equations~(\ref{eq:689}),~(\ref{eq:692}), and~(\ref{eq:695}) to Dirac's generators, see Appendix B. In addition, the commutation relations for the ten generators are given in Appendix B. The group Sp(4) has an important role in
squeezed state of light, polarization optics, and when
special relativity is incorporated with quantum mechanics~\cite{simon88,arvind95,chacon2021}.

The generators of this group are all defined
as four-by-four traceless matrices with only imaginary elements. As the group $Sp(4)$ contains a rotation subgroup in the form of the generators $J_i$ where $i=1,~2,3$, and two Lorentz subgroups (see Appendix A) in the form of the three boost generators $K_i$ with the rotation generators $J_i$, and the $Q_i$ with the rotation generators $J_i$, this group is isomorphic to $SO(3\, ,2)$. Thus, the $Sp(4)$ group is also the symmetry group for two Wigner phase spaces.

\section{Overlap of Wigner functions and Squeezed States of Light}\label{sec:92}

\begin{figure}[!ht] 
\centering
\begin{tikzpicture}


\draw[black,  thick, ->, >=stealth] (0,3) -- (6,3);
\draw[black,  thick,->, >=stealth] (3, -0.2) -- (3, 6.2); 
\path (6.2,3) node [black] {$x$};
\path (3,6.4) node [black] {$p$};

\draw[ultra thick, blue] (3,3) ellipse (2.5 cm and 0.4 cm); 

\draw[ultra thick,red, rotate around={90.0:(3,3)}] (3,3) ellipse (2.5 cm and 0.4 cm);

\draw[color=red, thick, variable=\an , domain=-60:240 , <-, >=stealth]
        plot ({5.2+0.3*sin(\an)}, {2+0.3*cos(\an)});
\path (4.6,2) node [red, scale=0.8] {Rotated};

\draw[black,  thick, ->, >=stealth] (7,3) -- (13.3,3);
\draw[black,  thick,->, >=stealth] (10.0, -0.2) -- (10.0, 6.2); 

\path (13.5,3) node [black] {$x$};
\path (10,6.4) node [black] {$p$};

\draw[ultra thick,red, rotate around={90.0:(10,3)}] (10,3) ellipse (2.5 cm and 0.4 cm);

\draw[ultra thick, violet] (10,5.1) ellipse (2.5 cm and 0.4 cm); 

\draw[thick, ->, >=stealth, color=violet] (13.0,3) -- node [fill=white,midway, scale=0.8, rotate=90] {Translated} (13.0,5.1);

\draw[color=red, thick, variable=\an , domain=-60:240 , <-, >=stealth]
        plot ({12.2+0.3*sin(\an)}, {2+0.3*cos(\an)});
\path (11.6,2) node [red, scale=0.8] {Rotated};     

\end{tikzpicture}
\caption{Overlapping Wigner functions.  On the left, the blue ellipse represents a squeezed state, while the red ellipse represents a rotated squeeze in the phase space. On the right, the red ellipse represents a rotated squeeze in the phase space, while the ellipse on the top is for the translated squeeze. The probability of the transition of one squeezed state into another is encoded in the overlap. It is seen that the overlapping regions differ depending on the type of the states involved.
}\label{fig:overlap}
\end{figure}
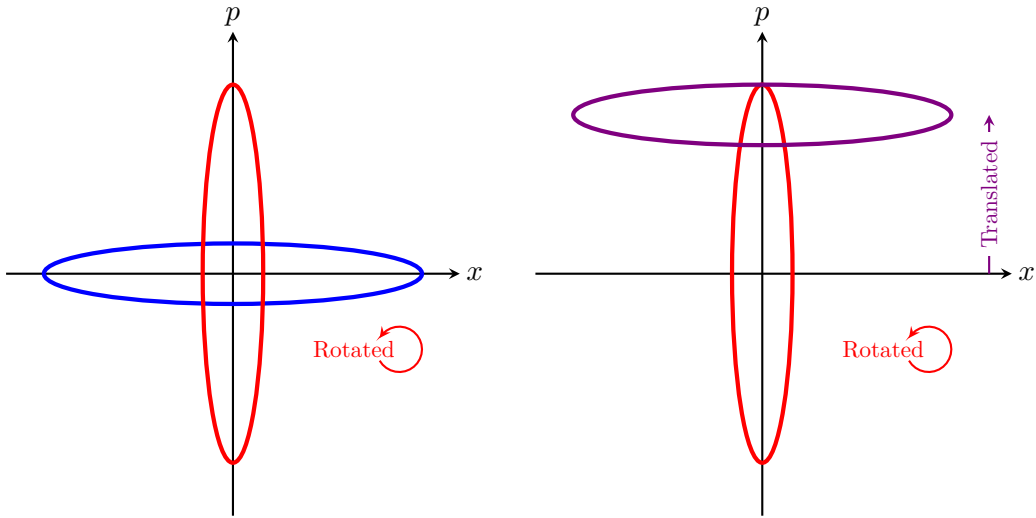

In quantum mechanics it is compelling to obtain measurable quantities. The quantity 
$|(\phi(x),\psi(x))|^{2}$ is directly measurable and is derived from the overlap of two different states. Here, 
$\phi (x)$ and $\psi (x)$ are wave functions for two different squeezed states.  As a simple case one can consider the transition probability of $\phi (x)$ to $\psi (x)$, where they are one-mode squeezed states. This can be expressed as
\begin{equation}
|\braket{\phi|\psi}| ^{2} = 2\pi \int 
W_{\phi}(x,p) W_{\psi}(x,p) \, ~dx ~dp \, .\label{eq:670a}    
\end{equation}
Since we are dealing with wave functions, the normalization constant is $2\pi$ for the integration of two Wigner functions.
From Equation~(\ref{eq:641}) we have for the ground state where $\lambda$ is the squeeze parameter
\begin{equation} 
W(0,-\lambda ;x,p) =  \left(  \frac{1}{\pi} \right)  \exp \left\{- \left(
e^{\lambda}  x^{2}  + e^{-\lambda}   p^{2} \right) \right\} \, .\label{eq:672} 
\end{equation} 
For the squeeze compressing the $p$-axis we have $W(0,\lambda ;x,p)$. This squeeze is in the opposite direction to the squeeze along the $x$-axis as the $x$-axis is elongated.
Hence the overlap becomes
\begin{eqnarray} \label{eq:723}
&&|\braket{0,-\lambda |0 ,\lambda }|^{2}  = \frac{2}{\pi} \int   \exp \left\{- \left(
e^{-\lambda}  x^{2}  + e^{\lambda}   p^{2} \right) \right\}  \exp \left\{- \left(
e^{\lambda}  x^{2}  + e^{-\lambda}   p^{2} \right) \right\}~dx~dp  \nonumber \\[1.0ex]
&&\hspace{24mm}= \frac{1}{\cosh \lambda} \, .
\end{eqnarray}
Two different types of overlap are depicted in Figure~\ref{fig:overlap}. The overlap depicted on the left of Figure~\ref{fig:overlap} results from Equation~(\ref{eq:723}). On the right of this figure we see a representation of an overlap between a translated squeeze and a rotated squeeze.
All squeezed pictures are initially elongated along the $x$-axis and contracted along the $p$-axis before any rotation or translation is applied. 

For two-mode squeezed states the overlap can be written as
$\braket{\beta ,\zeta_{2} | \alpha ,\zeta _{1} }$ 
where $\beta$ and $\alpha$ are two coherent states as defined in Equation~(\ref{co547}) and $\zeta_{1}, \zeta_{2}$ are defined as in Equation~(\ref{eq:631a}). 
The measurable quantity, as given in Equation~(\ref{eq:670a}), in terms of the corresponding Wigner functions is 
\begin{equation} 
|\braket{\beta  ,\zeta_{2} |\alpha  ,\zeta _{1}}| ^{2} = 2\pi \int
W(\beta   ,\zeta_{2} ;x_1,p_1) W(\alpha,\zeta_{1} ;x_2,p_2)~dx_1~dp_1~dx_2~dp_2 \, . \label{eq:670} 
\end{equation} 
This measures the transition probability, which literally comes from the overlap of distribution functions. 
Because of the canonical invariance, the integral in Equation~(\ref{eq:670}) can be brought to the form
\begin{equation} 
|\braket{0,-\lambda |\alpha ,\zeta }|^{2}  = 2\pi \int W(0,-\lambda;x_1,p_1) W(\alpha
,\zeta ;x_2,p_2)~dx_1~dp_1~dx_2~dp_2 \, .\label{eq:671} 
\end{equation} 
As for other properties of the two-mode state, the
overlap integral
of Equation~(\ref{eq:670a}) can be generalized to 
\begin{equation}
\hspace{5mm} |(\phi (x_{1},x_{2}), \psi (x_{1} ,x_{2})) |^{2}
= 2\pi  \int W _{\psi} (x_{1},p_{1};x_{2},p _{2}) 
W_{\phi} (x_{1},p_{1};x_{2},p_{2} )~dx_{1}~dp_{1}~dx_{2}~dp_{2} \, . \label{eq:697} 
\end{equation}

If the number operator is an addition of those for the first and second modes,
the generalization of the expectation value 
is straight-forward. For example,
\begin{equation} 
\braket{ \left(   \hat{N}_{1}  + \hat{N} _{2} \right) } = \braket{ \hat{N}_{1}} + \braket{\hat{N}_{2}} \, , \label{eq:698} 
\end{equation}
where
\begin{eqnarray}\label{eq:698a}
\braket{\hat N_{1}} & = & \braket{\left( J_{0}  +  J _{3}
 \right)}  = 
\frac{1}{2} \int \left( x_{1}^{2}  + p _{1}^{2} - 1  \right) W(x_{1} ,p_{1} ,x_{2} ,p_{2} )~dx_{1}~dp_{1} ~dx_{2}~dp _{2} \, , \nonumber \\[1.0ex]
\braket{\hat N_{2}} & = & \braket{\left(   J_{0}  - J_{3}  \right) }  = \frac{1}{2} \int \left( 
x_{2}^{2}  + p_{2}^{2}  - 1  \right)  W(x_{1} 
,p_{1},x_{2},p_{2} )~dx_{1}~dp_{1}~dx_{2}~dp_{2} \, . 
\end{eqnarray} 
Likewise, 
\begin{eqnarray} 
\braket{\hat N^{2}_{1}} & = & \frac{1}{4} \int \left[  \left(  x^{2}_{1}  + p^{2}_{1}-1
\right) ^{2}  -1 \right] W(x_{1} ,p_{1} ,x_{2} p_{2})~dx_{1}~dp_{1}~dx_{2}~dp_{2} \, , \nonumber \\[1.0ex]
\braket{\hat N^{2}_{2}} & = & \frac{1}{4} \int \left[  \left(  x_{2}^{2} +
p^{2}_{2} - 1 \right) ^{2} -1 \right]  W(x_{1},p_{1},x_{2},p_{2}
)~ dx_{1}~ dp_{1} ~dx _{2} ~dp _{2} \, .\label{eq:699} 
\end{eqnarray} 
As can be seen above, the integration is trivial over one of the modes. Then with the expression for the ground state Wigner function 
\begin{equation} \label{eq:700}
 W(x_1,p_1,x_2,p_2)=\frac{1}{\pi^{2}}\exp\left\{-\left(e^{-\lambda}x_{1}^{2} + e^{\lambda}p_{1}^{2} +e^{\lambda}x_{2}^{2} + e^{-\lambda}p_{2}^{2}\right)\right\}  \, , 
\end{equation}
we have
\begin{eqnarray}\label{eq:699BB}
& &\hspace{-2.5cm}\braket{(   \hat{N}_{1}  + \hat{N} _{2})} = 
 \frac{1}{2\pi^{2}}\int \left\{ \left( x_{1}^{2}  + p _{1}^{2} - 1  \right) + \left( 
x_{2}^{2}  + p_{2}^{2}  - 1  \right)\right \} \nonumber \\[2ex] & & \hspace{-0.4cm}\times \exp\left\{-\left(e^{-\lambda}x_{1}^{2} + e^{\lambda}p_{1}^{2} +e^{\lambda}x_{2}^{2} + e^{-\lambda}p_{2}^{2}\right)\right\} ~dx_{1} ~dp_{1} ~dx_{2} ~dp_{2} \nonumber \\
& &\hspace{-0.5cm} = (\cosh\lambda -1) \, .
\end{eqnarray}

Now, the expectation value of the product of two number operators is
\begin{equation} 
\braket{\hat N_{1} \hat N_{2}}  = \frac{1}{4} \int \left( x^{2}_{1}  + p_{1}^{2}  -
1 \right) \left(   x^{2}_{2}  + p^{2}_{2} - 1\right)
W(x_{1},p_{1},x_{2},p_{2} )~ dx_{1}  ~dp_{1} ~dx_{2} ~dp_{2} \, .\label{eq:6100} 
\end{equation}
This has the form of a correlation function of the fields, often interpreted as an intensity correlation between the two modes. Using again Equation~(\ref{eq:700}) we have
\begin{eqnarray}
& & \hspace{-2cm} \braket{\hat N_{1} \hat N_{2}}  = \frac{1}{4\pi^{2}} \int \left( x^{2}_{1}  + p_{1}^{2}  -1 \right) \left(   x^{2}_{2}  + p^{2}_{2} - 1\right) \nonumber  \\[2ex]
& & \hspace{-0.4cm} \times \exp\left\{-\left(e^{-\lambda}x_{1}^{2} + e^{\lambda}p_{1}^{2} +
e^{\lambda}x_{2}^{2} + e^{-\lambda}p_{2}^{2}\right)\right\}~dx_{1} ~dp_{1} ~dx_{2} ~dp_{2}
\,  \label{eq:6100b} 
\end{eqnarray}
yielding
\begin{eqnarray}
& & \hspace{-2cm}
\hspace{4mm} \braket{\hat N_{1} \hat N_{2}}  = \frac{1}{4\pi^{2}}\int \left( e^{\lambda}x^{2}_{1}  + e^{-\lambda}p_{1}^{2} -1 \right) 
 \left(  e^{-\lambda} x^{2}_{2}  + e^{\lambda}p^{2}_{2} - 1\right) \nonumber  \\[2ex]
& & \hspace{-0.2cm} \times\exp\left\{-\left(x_{1}^{2} + p_{1}^{2} +
x_{2}^{2} + p_{2}^{2}\right)\right\}~dx_{1} ~dp_{1}~dx_{2} ~ dp_{2} \nonumber \\
& & \hspace{-0.2cm}=\frac{1}{4}(\cosh{\lambda} -1)^2 \, .
\label{eq:6100c} 
\end{eqnarray}
This is in fact 
$\braket{\hat{N}_{1} \hat{N}_{2}}=\braket{\hat{N}_1}\braket{\hat{N}_2}$, meaning that the photon numbers are not correlated.  
 
It is also possible to calculate  $\braket{\left( \Delta \hat N_{1} \right) ^{2}}$  and $\braket{\left( \Delta \hat N_{2} \right)^{2}}$.
Due to the symmetry, we have with Equation~(\ref{eq:700}):
\begin{eqnarray} 
\braket{\hat N_{1}}=\frac{1}{2}\left(\cosh{\lambda}-1\right) \,\quad \mbox{and} \quad
\braket{\hat N_{1}^{2}}= \frac{1}{8} \left( 3\cosh(2\lambda) - 4\cosh(\lambda) + 1 \right) \, ,\label{eq:8n1}\\
\braket{\hat N_{2}}=\frac{1}{2}\left(\cosh{\lambda}-1\right) \,\quad \mbox{and} \quad
\braket{\hat N_{2}^{2}}= \frac{1}{8} \left( 3\cosh(2\lambda) - 4\cosh(\lambda) + 1 \right) \, \label{eq:8n2}.
\end{eqnarray}
The equations in Equation~(\ref{eq:8n1}) are the same as we saw in Equations~(\ref{eq:650}) and~(\ref{eq:652}), respectively.
This is due to the fact that the two-mode Wigner function in Equation~(\ref{eq:700}) can be factorized, i.e., the modes are 
not entangled. Thus, $\braket{\hat N_{1}^{2}}$ is entirely determined by the reduced state of the first mode, 
since integrating over $(x_2,p_2)$ has traced out the second mode. In more general terms, any measurement on the first mode
does not depend on an uncorrelated second mode.

By symmetry, the equations in  Equation~(\ref{eq:8n2}) are also the same. 
Then 
\begin{equation} 
\braket{(\Delta \hat N_{1})^{2}}=\frac{1}{2} \sinh^{2}\lambda  
\, , \label{eq:8delta}
\end{equation}
as we saw in Equation~(\ref{eq:653}).
Through similar calculations we obtain $\braket{(\Delta \hat N_{2})^{2}}= \braket{(\Delta \hat N_{1})^{2}}$.

\section{Two Coupled Oscillators} \label{sec:sqz06}
  
Using two harmonic oscillators 
the symmetries derived in Section~\ref{sec:68} will be examined.
The total Hamiltonian of two harmonic oscillators can be separated into
that of two independent systems if appropriate coordinate transformations are applied.  
As a result, all the
transformation matrices can become diagonal.
The Hamiltonian for each oscillator as given in Section~\ref{sec:4} using Equations~(\ref{eq:339}) and~(\ref{eq:340}) is 
\begin{equation}\label{sq602}
\mathcal{H}_{i} = \frac{1}{2}\left(x_{i}^{2} + p_{i}^2 \right) \, ,
\end{equation}
for $i=1,2$.
Then the total Hamiltonian can be written as
\begin{equation}
 \mathcal{H}_{+} = \mathcal{H}_{1} + \mathcal{H}_{2} = 
\frac{1}{2}\left(x_{1}^2 + x_{2}^2 + p_{1}^2 + p_{2}^2\right) \,
\end{equation}
We can also consider
\begin{equation}
\mathcal{H}_{-} = \mathcal{H}_{1} - \mathcal{H}_{2} = 
 \frac{1}{2}\left(x_{1}^2 - x_{2}^2 + p_{1}^2 - p_{2}^2\right) \, .\label{sq603}
\end{equation}
This Hamiltonian $\mathcal{H}_{-}$ produces the same set of wave functions as the
Hamiltonian $\mathcal{H}_{+}$.

While the Hamiltonian $\mathcal{H}_{+}$ is invariant under rotations
\begin{equation}\label{sq607}
\begin{pmatrix}x^{\prime}_{1} \cr x^{\prime}_{2}\end{pmatrix} = \begin{pmatrix}\;\;\;\cos\theta & \sin\theta
\cr -\sin\theta& \cos\theta\end{pmatrix}\begin{pmatrix}x_{1} \cr x_{2}\end{pmatrix} \, ,\qquad \begin{pmatrix}p^{\prime}_{1} \cr p^{\prime}_{2}\end{pmatrix} = \begin{pmatrix}\;\;\; \cos\theta & \sin\theta \cr
-\sin\theta & \cos\theta\end{pmatrix}\begin{pmatrix}p_{1} \cr p_{2}
\end{pmatrix} \, ,
\end{equation}
$\mathcal{H}_{-}$ is invariant under the squeeze transformations~\cite{fkr71} :
\begin{equation}\label{sq609}
\begin{pmatrix}x^{\prime}_{1} \cr x^{\prime}_{2}\end{pmatrix} = \begin{pmatrix}\cosh\lambda & \sinh\lambda
\cr \sinh\lambda & \cosh\lambda\end{pmatrix}\begin{pmatrix}x_{1} \cr x_{2}\end{pmatrix} \, ,
\qquad
 \begin{pmatrix}p^{\prime}_{1} \cr p^{\prime}_{2}\end{pmatrix} =
\begin{pmatrix} \cosh\lambda & \sinh\lambda \cr 
\sinh\lambda & \cosh\lambda\end{pmatrix}\begin{pmatrix}p_{1} \cr p_{2}\end{pmatrix}\,  ,
\end{equation}
where $\lambda$ and $\theta$ have been described in Section~\ref{sec:6}.

Now we consider two diagonal squeeze matrices $T_{0}$ and
$T_{1}$ applicable to the
four-component vector $\left(x_{1}, p_{1}, x_{2}, p_{2}\right)^T$.
The coordinate transformation matrices operating on the above vector are  \begin{equation}\label{sq601}
T_{0}(\lambda) = \begin{pmatrix}e^{\lambda} & 0 & 0 & 0 \cr
0 & e^{\lambda} & 0 & 0  \cr 0 & 0 &
e^{-\lambda} & 0 \cr 0 & 0 & 0 & e^{-\lambda} \end{pmatrix}\, ,
\quad \quad
T_{1}(\lambda) =  \begin{pmatrix}e^{\lambda} & 0 & 0 & 0 \cr
0 & e^{-\lambda} & 0 & 0  \cr 0 & 0 & e^{-\lambda} & 0 \cr 0 & 0 & 0 & e^{\lambda} 
\end{pmatrix} \, ,
\end{equation}
respectively. Referring to Section~\ref{sec:6}, $T_{0}$ performs
non-canonical transformations while $T_{1}$ performs
canonical transformations.

The ground-state Wigner function is given in Equation~(\ref{eq:688}).
Under the passive canonical transformation of
$T_{1}(\lambda)$, this Wigner 
function becomes:
\begin{equation} \label{eq:847}
W_{c,\lambda} = \left(\frac{1}{\pi}\right)^{2}
\exp\left\{-\left(e^{-2\lambda}x_{1}^{2} + e^{2\lambda}p_{1}^{2} +
e^{2\lambda}x_{2}^{2} + e^{-2\lambda}p_{2}^{2}\right)\right\} \, .
\end{equation}
The area of phase space is preserved for each mode.
As for the non-canonical transformation of $T_{0}(\lambda)$, the Wigner
function becomes~\cite{knp91,davies75,kn12symm}
\begin{equation}\label{eq:848}
W_{nc, \lambda} = \left(\frac{1}{\pi}\right)^{2}\exp\left\{
-\left(e^{-2\lambda}x_{1}^{2} + e^{-2\lambda}p_{1}^{2} +
e^{2\lambda}x_{2}^{2} + e^{2\lambda}p_{2}^{2}\right)\right\} \, .
\end{equation}
The area of the first
mode shrinks while the other expands. The product of these two areas
remains constant. In both Wigner equations the phase spaces are entangled unless $\lambda =0$. Recalling Equation~(\ref{eq:312}) that the Wigner function cannot be positive everywhere in phase space, there is an interesting article~\cite{rahman_local_2025} which discusses the entanglement of two interacting qubits in phase space using the Wigner function negativity, continuous variable, and non-classicality. The article proposes an experiment to verify the method and results.  Entanglement itself has been experimentally verified in the 1960's and an article describing this experiment in detail, by the experimenter, is given in~\cite{kocher_quantum_2024}.

\subsection{Density Matrix for Two Harmonic Oscillator States}\label{ent03}

The density matrix for the two harmonic oscillator is examined now.
The ground-state wave function of two coordinate
variables can take the form
\begin{equation}\label{en301}
 \psi_{0,0}(x_{1},x_{2}) = \frac{1}{\sqrt{\pi}}
  \exp{\left\{-\frac{1}{2}\left(x_{1}^2 + x_{2}^2\right)\right\}} \, ,
\end{equation}
where this wave function corresponds to the vacuum state of two different
photons.  This function is still separable in the $x_{1}$- and
$x_{2}$-coordinates, and thus is not entangled. 

Now the coordinates will be rotated and
squeezed.  Under the coordinate rotation
\begin{equation}\label{en303}
 \begin{pmatrix}x_{r1} \cr x_{r2}\end{pmatrix} = \begin{pmatrix} \,\,\cos\theta & -\sin\theta \cr \, \sin\theta & \;\;\;\cos\theta\end{pmatrix}  
\begin{pmatrix}x_{1} \cr x_{2} \end{pmatrix} \, ,
\end{equation}
the wave function remains separable and
is now written as
\begin{equation}\label{en304}
\psi_{0,0}(x_{1},x_{2}) = \frac{1}{\sqrt{\pi}}
  \exp\left\{-\frac{1}{2}\left[\left(x_{1}\cos\theta +
    x_{2}\sin\theta\right)^{2} + 
  \left(-x_{1}\sin\theta + x_{2}\cos\theta\right)^{2}\right]\right\}.
\end{equation}
The squeeze transformation acts on the rotated coordinates as:
\begin{equation}\label{en305}
 \begin{pmatrix}x_{s1} \cr x_{s2}\end{pmatrix} =
 \begin{pmatrix}e^{\lambda} & 0 \cr 0 & e^{-\lambda}\end{pmatrix} 
 \begin{pmatrix}x_{r1} \cr x_{r2}
 \end{pmatrix} \, .
\end{equation}
Carrying out these transformations,
the wave function becomes
\begin{equation} \label{en307}
\hspace{5mm} 
\psi_{\theta,\lambda}\left(x_{1}, x_{2}\right) = \frac{1}{\sqrt{\pi}}
\exp\left \{  - \frac{1}{2}\left [ e^{-2\lambda}\left(x_{1}\cos\theta +
  x_{2}\sin\theta\right)^{2}  
+ e^{2\lambda}\left(-x_1\sin\theta +x_2\cos\theta \right)^{2} \right ] \right\}.
\end{equation} 
This wave function remains separable in the $x_{1}$ and $x_{2}$
variables unless both the $\theta$
and $\lambda$ variables are non-zero as
indicated in Table~\ref{entab11}.

The density matrix is defined as
\begin{equation} \label{en309}
\rho_{\theta,\lambda}\left(x_{1},x_{2};x^{\prime}_{1},x^{\prime}_{2} \right)
=\psi_{\theta,\lambda}\left(x_{1},x_{2}\right)
\psi^{*}_{\theta,\lambda}\left(x^{\prime}_{1},x^{\prime}_{2}\right) \, ,
\end{equation}
when all variables are measured~\cite{feynman_2019}.
This can be written as
\begin{eqnarray}\label{en311}
&{}& \hspace{-12mm}\rho_{\theta,\lambda}\left(x_{1},x_{2};x^{\prime}_{1},x^{\prime}_{2} \right) = \frac{1}{\pi}
 \exp\left \{ -\frac{1}{2} e^{-2\lambda}\left[(x_{1}\cos\theta + x_{2}\sin\theta)^2
 + (x^{\prime}_{1}\cos\theta + x^{\prime}_{2}\sin\theta)^2 \right]\right\}  \nonumber \\[2mm]
&{}& \hspace{10mm}\times \exp\left\{-\frac{1}{2} e^{2\lambda}\left[(-x_{1}\sin\theta
+ x_{2}\cos\theta)^2 + (-x^{\prime}_{1}\sin\theta + x^{\prime}_{2}\cos\theta)^2
\right]\right\}.
\end{eqnarray} 
This expression is consistent with the condition 
\begin{equation}\label{en312}
Tr\left(\rho_{\theta,\lambda}\right) =
Tr\left(\rho^{2}_{\theta,\lambda}\right) = 1
\end{equation}
for pure states, where
\begin{equation} \label{en309a}
Tr\left(\rho_{\theta,\lambda}\right) =\int\rho_{\theta,\lambda}\left(x_{1},x_{2};x_{1},x_{2}\right)\, dx_{1}\,dx_{2} \, .
\end{equation}
If no observations are made on the $x_{2}$ variable
the density matrix for $x_{1}$ is
\begin{equation}\label{en315}
 \rho(x,x^{\prime}) = \int \rho(x, x_{2}; x^{\prime}, x_{2})~dx_{2} \, ,
\end{equation}
where $x_{1}$ is replaced with $x$.  The evaluation of this integral
leads to
\begin{eqnarray}\label{en317}
&{}& \rho(x, x^{\prime}) = \left[\frac{1} {\pi (\cosh(2\lambda) -
\sinh(2\lambda)\cos(2\theta))} \right]^{1/2} \nonumber\\[2mm]
&{}& \hspace{5mm}\times \exp\left\{- \left[\frac{(x + x^{\prime})^{2} +
(x - x^{\prime})^{2}(\cosh^{2}(2\lambda) - \sinh^{2}(2\lambda) \cos^{2}(2\theta) )}
{4(\cosh(2\lambda) - \sinh(2\lambda)\cos(2\theta))}\right]\right\} \, .
\end{eqnarray}

\begin{table}
\caption{Two entangled oscillators.  Unless both parameters are non-zero,
the variables $x_{1}$ and $x_{2}$ remain separable.}\label{entab11}
\vspace{-1mm}
\begin{center}
\begin{tabular}{ccccc}
\hline\hline \\[-0.9ex]
\hspace{10mm} &\hspace{15mm}& $\theta = 0$ & \mbox{}\hspace{20mm} \mbox{}& $\theta \neq 0 $ \\
\hline\\
$\lambda = 0$ &\mbox{}& Separable  &\mbox{}& Separable \\[3mm]
\hline \\
$\lambda \neq 0 $  &\mbox{}& Separable &\mbox{}& ENTANGLED  \\[3mm]
\hline\hline\\[-0.8ex]
\end{tabular}
\end{center}

\end{table}
With this expression
the trace integral for this reduced state obtained by integrating over $x_2$ is given by
\begin{equation}\label{en319}
Tr(\rho) = \int \rho(x,x)~dx
\end{equation}
becomes one, as it should be for all density matrices.  As for $Tr(\rho^{2})$,
the result of the trace integral becomes
\begin{equation}\label{en321}
Tr(\rho^{2}) = \int 
\left \{\rho(x,x^{\prime}) \rho(x^{\prime},x)~dx^{\prime}\right \} ~dx =
\frac{1}{\sqrt{1 + \sinh^{2}(2\lambda) \sin^{2}(2\theta)}} \, .
\end{equation}
This is less than one for non-zero values of $\lambda$ and $\theta$. This is
consistent with the general theory of density matrices.  If
$\lambda = 0$ and/or $\theta = 0$, the first oscillator is totally
independent of the second oscillator, and the system of the first
oscillator is in a pure state, and $Tr(\rho^{2})$ becomes one.  
However, according to Table~\ref{entab11}, the density matrix and therefore the
harmonic oscillators become entangled when
both $\lambda$ and $\theta$ are non-zero and $Tr(\rho^{2})$
is also less than one. 

We noted earlier that the density matrix is especially important when not all 
measurable variables are measured in laboratories. However,
the space in which we cannot make measurements is entangled with the space in which we make measurements. It is possible to interpret this 
second space, which obviously has an effect on the first, as Feynman's "rest of the universe"~\cite{hkn99ajp}. This effect is taken care of by partially integrating over the variables of the second space.  
The idea of Feynman's rest of the universe also presents itself in Section~\ref{sec:14.3}, where we discuss the coupling of two concentric Poincaré spheres.  These ideas often emerge in "open quantum systems" and are highly instrumental in measurement theory~\cite{breuer_2010}.

\section{Density Matrix and the Poincaré Sphere}\label{sec:14.3}  

Among his many foundational contributions to mathematics and physics, Henri Poincaré introduced a geometric representation of the polarization states of light by mapping them to points on a sphere; now known as the Poincaré sphere~\cite{hkn97}.
In the conventional optical framework, three of the four Stokes parameters correspond to points on the surface of the Poincaré sphere, while the fourth parameter represents the total intensity.  In the Lorentzian regime, it is shown that these four parameters transform as a four-vector, with the fourth parameter being the timelike component. The Stokes parameters are directly related to the coherency matrix, which in quantum mechanical terms is equivalent to the density matrix~\cite{born99}. The off-diagonal elements of this matrix represent the degree of coherence (or decoherence) between orthogonal polarization or between quantum states.

Transformations under the two-by-two Lorentz group preserve the determinant of the coherency (or density) matrix, and hence leave the decoherence parameter invariant. In this section we shall address how the issue of decoherence can be treated as a symmetry problem. 

\subsection{Stokes vectors, coherency matrix and the Poincaré sphere}

We first dwell on the formulation of coherency between the two orthogonal electric fields:
\begin{equation}\label{eq:901}
E_{x}=A\,\exp{i(kz-wt+\phi_{1})} \qquad \mbox{and} \qquad E_{y}=B\,\exp{i(kz-wt+\phi_{2})} \, ,
\end{equation}
where $z$ is the propagation direction, $A$ and $B$ are the amplitudes of the field along 
the $x$ and $y$ directions, respectively. The angle $\phi=\phi_{1}-\phi_{2}$ is the phase difference
between the components.

The elements of the coherency
matrix can be associated as~\cite{born99,bross98}: 
\begin{eqnarray}\label{sto304}
&{}& S_{11} = <E_{x}^{*}E_{x}> , \qquad S_{22} = <E_{y}^{*}E_{y}> ,
\nonumber \\[1ex] 
&{}& S_{12} = <E_{x}^{*}E_{y}> ,  \qquad
S_{21} = <E_{y}^{*}E_{x}> \, ,
\end{eqnarray}
where
\begin{equation}\label{sto301}
C = \begin{pmatrix}
S_{11} & S_{12} \cr S_{21} & S_{22}
\end{pmatrix}
\end{equation}
is the form of the matrix.
Here for some field $\psi_{i}$, $<...>$ is defined as
\begin{equation}
<\psi^{*}_{i}\,\psi_{j}>=\frac{1}{T}\int^{T}_{0} \psi^{*}_{i}(t^{\prime})\psi_{j}(t+t^{\prime})\, dt^{\prime}
\end{equation}
to be the time average.
The absolute values of $E_x$ and $E_y$,
respectively, are given by the diagonal elements.  If the two
transverse components are not completely 
coherent, the off-diagonal elements could be smaller than the product
of $E_x$ and $E_y$.  The degree
of decoherence in the system is specified by
$\tau t$ with $\tau>0$. The decoherence is minimum if $t$ is zero,
and becomes maximum if $t$ attains very large values.
We then have the  coherency matrix as~\cite{hkn97josa}: 
\begin{equation}\label{sto305}
C = \begin{pmatrix} A^2 & AB~e^{-(\tau t + i\phi)} \cr
AB~e^{-(\tau t - i\phi)} & B^2 
\end{pmatrix} \, .
\end{equation}
The transformation matrices applicable to the Jones vector,
derived from Equation~(\ref{eq:901}), are the two-by-two representations of the Lorentz
group~\cite{bkn19iop}. In this section we are particularly interested in the transformation
matrices as applicable to the coherency matrix.
\par
The determinant of the above coherency matrix is calculated to be
\begin{equation}\label{sto306}
\det(C) = (AB)^2 \left(1 - e^{-2\tau t}\right), 
\end{equation}
and its trace is obtained as
\begin{equation}\label{sto306a}
\mbox{tr}(C) = A^2 + B^2 .
\end{equation}

From the components of the coherency matrix, it will prove to be useful to 
introduce four quantities as
\begin{eqnarray}\label{sto312}
&{}& S_{0} = \frac{1}{2}(S_{11} + S_{22}),  \qquad
    S_{3} =  \frac{1}{2}(S_{11} - S_{22}),  \nonumber \\[2ex]
&{}& S_{1} = \frac{1}{2}(S_{12} + S_{21}), \qquad
S_{2} = \frac{i}{2}(S_{12} - S_{21}).
\end{eqnarray}
They are called Stokes parameters, 
and the four-by-four transformation matrices applicable to these parameters are 
widely known as Mueller matrices~\cite{mueller43,azzam77,bross98}. These matrices
are responsible for the relationship between polarization states of the incident 
light and the emerging light after passing through any number and any type of optical 
elements, such as polarizers, waveplates, and scatterers.  It is by employing the Mueller 
matrices that we can perform Lorentz transformations on 
the four Stokes parameters. Indeed, Stokes parameters behave as 
the components of a four-vector~\cite{hkn97}.  

Now, using Equation~(\ref{sto305}) and  Equation~(\ref{sto312}), the components of this four vector are found to be:
\begin{eqnarray}\label{4vec33}
&{}& S_{0} = \frac{1}{2}  (A^2 + B^2) \, , \qquad \hspace{3mm} S_{1} = AB (\cos\phi)e^{-\tau t} \, ,\nonumber \\
&{}&S_{2} = AB(\sin\phi)e^{-\tau t} \, ,  \qquad S_{3} = \frac{1}{2} (A^2 - B^2) \,.
\end{eqnarray}
Conversely, the coherency matrix can succinctly be 
constructed from Stokes parameters and the Pauli spin matrices as 
\begin{equation}\label{sumsto}
C=\sum_{i=0}^{3} \sigma_{i}\,S_{i} \, , 
\end{equation}
where $\sigma_{0}$ is the two-by-two identity matrix.
There is a close relation between the coherency matrix $C$ and the density matrix~\cite{blum_2012}:
\begin{equation}
    \rho \;=\; \frac{1}{2} C
\end{equation}
which henceforth, provides the relation between the Stokes parameters and the density matrix.
It is also well-known that the Stokes vectors can be mapped to the Poincaré sphere.

Originally the Poincaré sphere proposed by Henri Poincaré was a three dimensional 
object for representing the polarization state of light.  Three of the Stokes vectors $S_{i}=
\{S_{1},\,S_{2},\,S_{3}\}$ correspond to Cartesian coordinates.
The radius of the Poincaré  sphere is given in terms of these three Stokes vectors as:
\begin{equation}\label{eq:1431}
	r = \sqrt{S_{1}^2 + S_{2}^2 + S_{3}^2} \, .
\end{equation}
While $S_0$ is known to represent light intensity, now can be considered to be the time-like 
component of a four-vector.   
We shall use this property of the Stokes 
vector to devise a second concentric sphere, the radius of  which is taken to be  $s=S_{0}$. 
Within the framework of the Lorentz group the relation between these two radii, namely $s^2-r^2$, 
is an invariant, and hence is the determinant of the density matrix.

\subsection{Two Concentric Poincaré Spheres}\label{sec:14.3-1}

We noted that the determinant of the density matrix is invariant in the Lorentzian regime and cannot tackle the decoherence process with the decaying 
of the off-diagonal components. As a result, the question arises as to whether this process can be reformulated by incorporating the symmetry properties of a larger Lorentz group. In the following, we discuss and address this issue in detail.

In view of  Equation~(\ref{4vec33}) and Equation~(\ref{eq:1431})  the radius $r$ of 
the sphere is rewritten as
\begin{equation}\label{eq:1432}
	r =\frac{1}{2} \sqrt{\left( A^2 -B^2\right)^2 + 4 (AB)^2 e^{-2\tau t} } 
\end{equation}
which describes the conventional Poincaré sphere.  However, the four-vector in 
Equation~(\ref{4vec33}), in addition to the space-like components, has a time-like component 
which provides another radius
\begin{equation}\label{eq:1433}
	s =\frac{1}{2} \left(A^2 + B^2\right)  
\end{equation} 
defining an outer sphere concentric to the inner sphere.  The quantity $s^{2}-r^{2}$ is Lorentz-invariant and is 
equal to the  value of the determinant in Equation~(\ref{sto305}).  The $t$ parameter cannot be 
changed in the Lorentzian regime, therefore cannot deal with the decoherence process by the decaying 
of the off-diagonal components.  However, this restriction can be  released, when the symmetry features of a larger Lorentz group is introduced,
particularly the $SO(3\,, 2)$ group. 

Now when $t=0$, the inner radius is equal to the outer radius. When $t$ is sufficiently large 
we have
\begin{equation}\label{eq:1434}
	S_{3}=\frac{1}{2} \left(A^2 - B^2\right) \, .
\end{equation}

Let us introduce a spherical coordinate system and identify the components of $S_{i}$ with those as 
\begin{eqnarray}\label{eq:1435}
	&&{} r_{z} =  r (\cos\theta)= (A^2 - B^2)/2  \, , \nonumber \\[2ex]
	&&{} r_{x} = 
	r (\sin\theta) \cos\phi =  AB (\cos\phi)e^{-\tau t}\,  , \nonumber \\[2ex]
	&&{} r_{y} = r (\sin\theta) \sin\phi =  AB (\sin\phi)e^{-\tau t} \,  .
\end{eqnarray}
Since, the Lorentz symmetry allows rotations in this three-dimensional scheme, 
with an appropriate rotation the four-vector can be brought to 
\begin{equation}\label{eq:1436}
    (s, \,r , \, 0 ,\, 0 )^{T} \, .
\end{equation}
The rotations do not change the radii of the outer and inner spheres, thus 
$s$ and $r$ remain invariant.  On the other hand, if  the above four-vector is boosted we have
\begin{equation}\label{eq:1438}
	\begin{pmatrix}\cosh \lambda & -\sinh \lambda & 0 & 0 \\
		-\sinh \lambda & \cosh \lambda &0 & 0 \\
		0&0&1&0\\
		0&0&0&1
	\end{pmatrix}\, \begin{pmatrix} s \cr r \cr 0 \cr 0 \end{pmatrix} = \begin{pmatrix}s(\cosh\lambda) - r (\sinh\lambda) \cr r (\cosh\lambda)
		- s (\sinh\lambda) \cr 0 \cr 0   \end{pmatrix} \, .
\end{equation}
We see that, this transformation changes the outer and inner radii, but
keeps $(s^2 - r^2)$ invariant.
One can choose the value of $\lambda$ such that~\cite{bk06}:
\begin{equation}\label{eq:1439}
	r \,(\cosh\lambda)- s \, (\sinh\lambda) = 0  \,  ,
\end{equation}
which leads to $\tanh\lambda = r/s$.   Then, using Equations~(\ref{eq:1432}) and~(\ref{eq:1433}), 
the four-vector on the right hand side of Equation~(\ref{eq:1438}) becomes
\begin{equation}\label{eq:14310}
	(\sqrt{s^2 - r^2}\, ,0,0,0 )^{T} =
	(AB \sqrt{1 - e^{-2\tau t}}\,,0,0,0 )^{T} \, .
\end{equation} 
Lorentz transformations bring the Poincaré sphere
to a one-number system.   

\begin{figure}[!t]
\begin{center}
\includegraphics[scale=0.2]{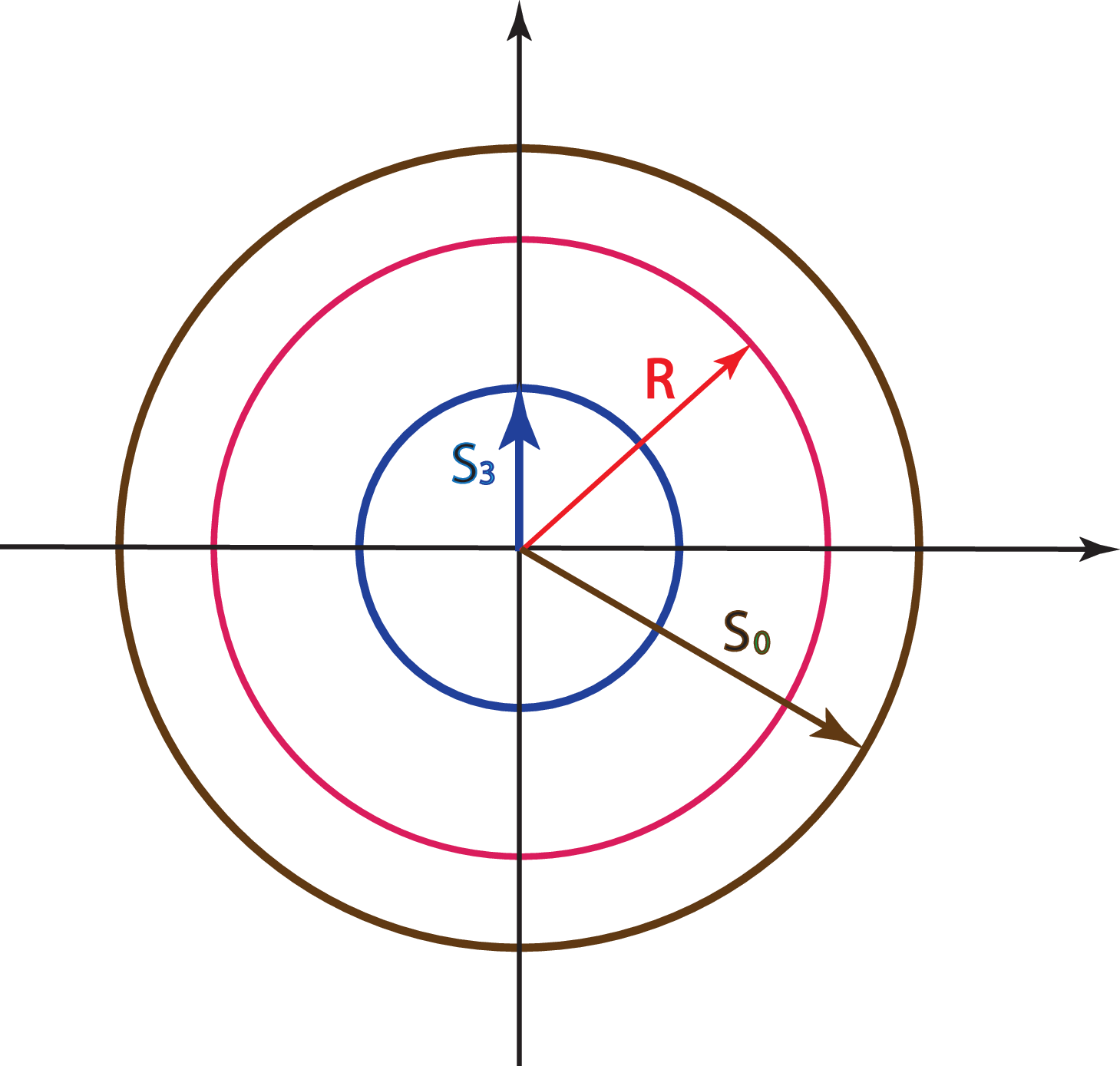} \hspace{2cm}
\includegraphics[scale=0.2]{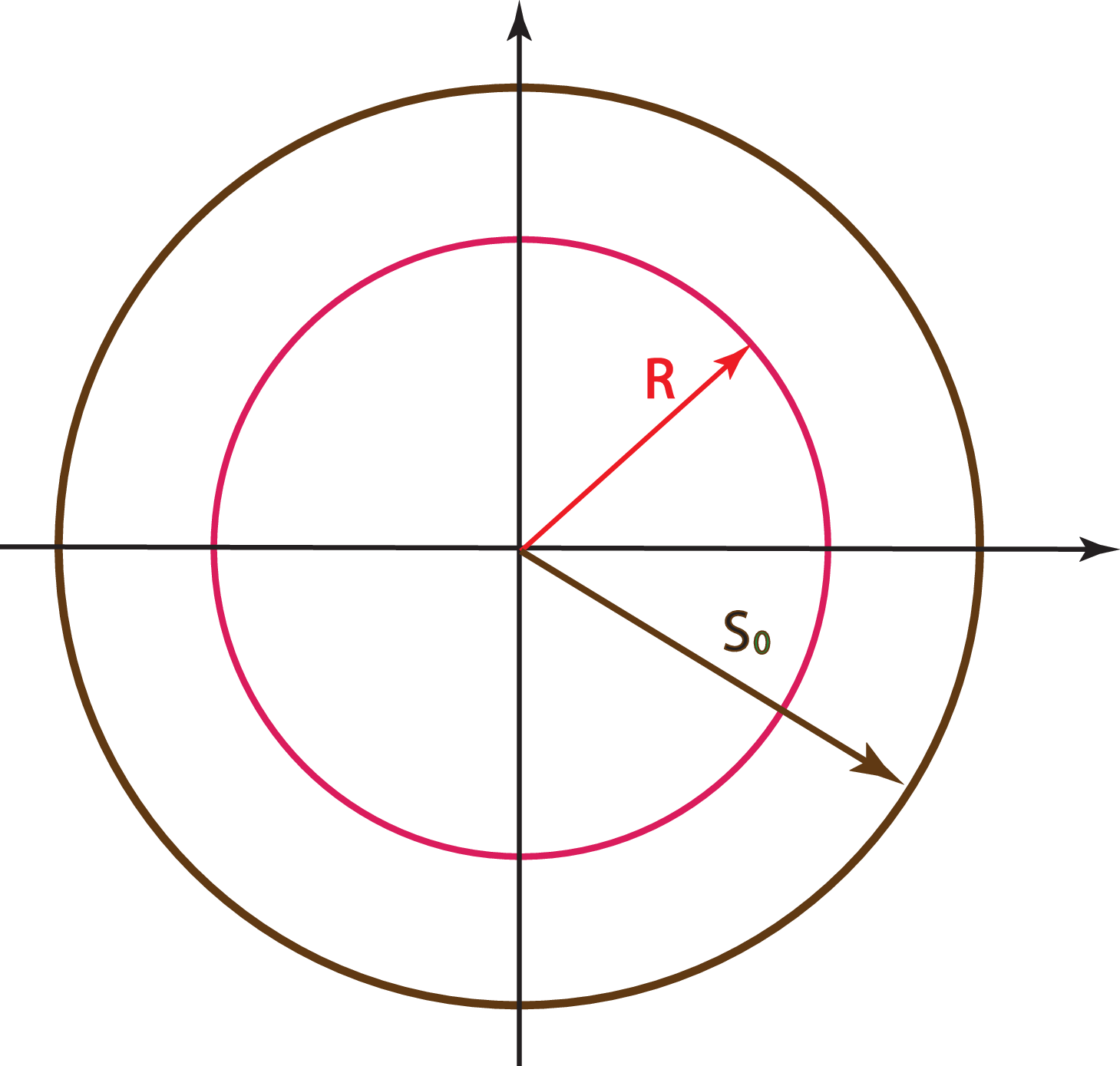}
\end{center}
\caption{Variable radius of  the Poincaré sphere.  From Equation~(\ref{eq:1433}) and Equation~(\ref{eq:14315}), the variable radius $R$ takes its maximum value $S_0$, 
when $\chi=0^{\circ}$.  It becomes minimum when the decoherency angle reaches 
$90^{\circ}$.  
Its minimum value is $S_3$ as is illustrated in the left figure.  The degree of polarization is maximum when $R = S_0$, and is minimum when $R = S_3$.  
According to Equation~(\ref{eq:1434}), $S_3$ becomes $0$ when $A= B$, thus the minimum value
of $R$ becomes zero, as is indicated in the right figure.  Its maximum
value is still $S_0$~\cite{kn13sym}.}\label{fig:1431}
\end{figure}

Now we shall see that it is possible to change the value of
$\left(s^2 - r^2\right)$ in the above expression by changing the
time variable $t$. This can be done by considering the rotation matrix between the
two time-like coordinates  
$(t_1,t_2,z,x,y)$ of  $SO(3\,,2)$.  Let us consider the transformation
of a five-vector by this rotation
\begin{equation}\label{eq:14311}
	\begin{pmatrix}\cos \chi & -\sin \chi & 0 & 0  &0\\
		\sin \chi & \cos \chi &0 & 0 & 0\\
		0&0&1&0&0\\
		0&0&0&1&0\\
		0&0&0&0&1
	\end{pmatrix}\begin{pmatrix}
	   v\cr0\cr0\cr0\cr0 
	\end{pmatrix}=\begin{pmatrix}
	    v \cos \chi \cr v \sin \chi \cr 0 \cr 0 \cr 0
	\end{pmatrix} \, .
\end{equation}	
The invariant quantity in this larger regime is $v^2$.  The $SO(3\,,2)$  group has two Lorentzian subgroups, where their respective four-vectors are expressed as
\begin{equation}\label{eq:14312}
(v \cos \chi \,, 0,0,0)^{T} \qquad \mbox{and} \qquad ( v \sin \chi \, ,0,0,0)^{T} \, .
\end{equation}
We compare the four-vector in Equation~(\ref{eq:14310}) with Equation~(\ref{eq:14312})  above and identify
the non-zero components.  Then we have 
\begin{equation}\label{eq:14313}
\left(s^2 - r^2\right)=v^{2 }\cos ^{2 }\chi \qquad \mbox{or} \qquad	\left(s^2 - r^2\right)= v^{2 }\sin ^{2}\chi
\end{equation}
which allows variable radii for the Poincaré spheres as illustrated in 
Figure~\ref{fig:1431}.  So, we can consider the relations
\begin{equation}\label{eq:14315}
	\cos \chi =e ^{-\tau t} \qquad \mbox{or} \qquad \sin\chi =e ^{-\tau t} \, ,
\end{equation}
by the redefinition of $ab$ as $v$ and since we have $\left(s^2 - r^2\right)=(AB)^{2} (1 - e^{-2\tau t})$. 

Let us now examine the diagonalization of the density matrices. For this purpose
the amplitudes are redefined as
\begin{equation}\label{eq:14215}
	A=\sqrt{2q}\,\cos(\delta/2)	\qquad  \mbox{or}  	\qquad  B=\sqrt{2q}\,\sin(\delta/2) \, .
\end{equation} 
Thus the density matrix  $\rho_{c}$ is rewritten 
\begin{equation}\label{eq:14216}
	\hspace{-0.2cm} \rho_{c}(\chi)=2q \begin{pmatrix}
		\cos^{2} (\delta/2)  & \sin (\delta/2) \cos(\delta/2) e^{-i\phi}(\cos\chi)  \\
		\sin (\delta/2) \cos(\delta/2) e^{i\phi}(\cos\chi)	 & 	\sin^{2} (\delta/2)   \end{pmatrix} \, .
\end{equation}
Following a similar procedure for $\rho_s$, they  become
\begin{equation}\label{eq:14217}
	\rho_{c}(\chi) = \frac{1}{2}\begin{pmatrix} 1 & e^{- i\phi}\cos\chi \cr  e^{i\phi}\cos\chi & 1  \end{pmatrix}  \qquad \mbox{or} \qquad
    \rho_{s}(\chi) = \frac{1}{2}\begin{pmatrix} 1 & e^{- i\phi}\sin\chi \cr  e^{i\phi}\sin\chi & 1  \end{pmatrix}  
\end{equation}
when the angle $\delta$ is conveniently chosen to be $90^{\circ}$.  In addition, the normalization condition $Tr(\rho)=1$ is maintained.   These can now be diagonalized to take the forms
\begin{equation}\label{eq:14218a}
	\rho_c(\chi) = \frac{1}{2}\begin{pmatrix} 1+\cos\chi  & 0 \\  0 & 1- \cos\chi  \end{pmatrix} 
\qquad \mbox{or} \qquad
	\rho_s(\chi) = \frac{1}{2}\begin{pmatrix} 1+\sin\chi  & 0 \\  0 & 1- \sin\chi   \end{pmatrix} \, .
\end{equation}

\FloatBarrier

The expressions in these forms give a better interpretation for the entropy in their respective domains.
von Neumann defined the entropy by~\cite{neum2018}:
\begin{equation}\label{eq:14219}
	S=-Tr(\rho\, \ln \rho)  \, .
\end{equation}
For the density matrix $\rho_c$,  entropy  becomes~\cite{kiwi90pl,eisert10,kn14}:
\begin{equation}\label{eq:14220}
	S_c(\chi) = - \frac{1 + \cos\chi}{2}\ln\left(\frac{1 + \cos\chi}{2}\right)
	-\frac{1 - \cos\chi}{2}\ln\left(\frac{1 - \cos\chi}{2}\right)  \, .
\end{equation}
The entropy $S_c$ of this space is a monotonically increasing function of $\chi$.  We also 
see that if the entropy is zero the system is
completely coherent with $\chi=0^{\circ}$.  When the system is totally
incoherent with $\chi = 90^{\circ}$ the 
entropy takes the maximum value of $\ln(2)$.

In a similar manner, the entropy of the second space can be expressed as
\begin{equation}\label{eq:14222}
	S_s(\chi) = - \frac{1 + \sin\chi}{2}\ln\left(\frac{1 + \sin\chi}{2}\right)
	-\frac{1 - \sin\chi}{2}\ln\left(\frac{1 - \sin\chi}{2}\right) .
\end{equation}
The entropy  $S_s$  of this second space, unlike the first one,  is a decreasing function of  $\chi$.  
The first of those belongs to the part of the universe in which we make measurements, while the second is in the "rest of the universe", as was discussed by von Neumann himself in~\cite{fey72}. The increase of entropy in the first space can be interpreted as the loss of energy in the second.  There is a decoherence-recoherence process unfolding in these coupled universes.  Yet, the sum of the entropies is not a conserved quantity of the total system.  This should not be problematic as the second space may not be a physical space, but falls into the description of von Neumann's "rest of the universe". 
On the other hand we can note that the sum of the determinants of the density matrices, 
$\det{\rho_{c}}+\det{\rho_{s}}$, is independent of the decoherency angle $\chi$ and thus in that respect, is a conserved quantity.

It is worthwhile to mention that in quantum mechanics and computing, there is an analogous sphere known as the Bloch sphere, a geometrical visualization of a pure state space of a two-level quantum mechanical system. In that context states are replaced by qubits, which are the basic units of quantum information~\cite{nielsen2010}. Quantum states of the Bloch sphere are not necessarily pure. In realistic settings, pure states eventually interact with the environment leading to the emergence of mixed states.
Mixed quantum states are located interior of the unit sphere, meaning that they are contained within the Bloch sphere.
Therefore, the problem of a variable radius also arises in this realm.
From a mathematical point of view, the normalization condition of the states can be relaxed to generate mixed states.  
 
\section{Conclusions}\label{sec:conc}

In this paper we have started by defining the classical and quantum phase space.  
Canonical transformations were then presented, after which the Wigner distribution function and the associated density matrix were discussed, followed by the harmonic oscillator in the Wigner phase space. Coherent and single mode squeezed states of light and the squeezed vacuum were studied using the Wigner functions as well as the form of the Wigner function
from which the squeezed vacuum is derived was discussed using the Wigner phase space. Then the symmetries of two-mode states of light were elucidated, and it was shown that the four dimensional Wigner function was required. Our treatment suggests that the Wigner phase space representation of
quantum mechanics can provide a more thorough examination of the two-mode squeezed state, thereby rendering the associated symmetry properties more transparent. We then employed the overlap of the Wigner distribution function in this four dimensional space to study two-mode squeezed states of light.  Two coupled harmonic oscillators including the way in which they become entangled as well as the density matrix associated with these coupled harmonic oscillators were studied. As an example of the use of density matrix we presented the Poincaré sphere with the Stokes parameters forming a four-vector. It was shown that a larger symmetry group, namely 
$SO(3,2)$, is required, since Lorentz invariance in the $SO(3,1)$ framework precludes any change in decoherence, whereas the description of the Poincaré sphere derived from the Stokes vectors inherently involves a changing degree of decoherence.
In this work, we made extensive use of Lie groups $E(2)$, $Sp(2)$, $ISp(2)$, $Sp(4)$, $SO(3,1)$, $SO(3,2)$ and subgroups, to first characterize and then exploit the symmetry properties of the Wigner functions and the density matrix. 
Although the results presented in this article are not new, they were drawn from earlier collaborative works by the authors and presented from a fresh point of view.

\appendix
\section[Appendix Title]{The SO(3,1) Lorentz Group} 
\numberwithin{equation}{section}
\setcounter{equation}{0}
The Lorentz group, $SO(3\, ,1)$ is a six-parameter Lie group. The generators of the group
consist of  the three rotation generators,  $J_1$, $J_2$, and $J_3$ that generate rotations about
the $x$-, $y$-, and $z$-axis, respectively. These generators form the familiar rotation group and are a
subgroup of the $SO(3\, ,1)$. The three boost generators are $K_1$, $K_2$, and $K_3$ which generate boosts
along the $x$-, $y$-, and $z$-axis, respectively. 
They form six independent generators for the Lie algebra of the Lorentz 
group~\cite{carmeli76}.
The commutation relations for these six generators form a closed set:
\begin{equation} 
[J_{i}, J_{j}] = i\varepsilon_{ijk} J_{k} \, , \qquad
 [J_{i}, K_{j}] = i\varepsilon_{ijk}  K_{k} \, , \qquad
[K_{i}, K_{j}] = -i\varepsilon_{ijk} J_{k} \, .
\label{eq:52}
\end{equation}
Thus they form the Lie algebra for the $SO(3\, ,1)$. 

We must also consider $SL(2\, ,C)$ that is the covering group of the Lorentz group and is
generated by:
\begin{equation} \label{eq:53}
J_{i} = (1/2)\sigma_{i} \qquad {\rm and} \qquad K_{i} = (i/2)\sigma_{i},
\end{equation} 
where $\sigma_{i}$ are the Pauli spin matrices defined as
\begin{equation}\label{eq:54}
\sigma_{1} = \begin{pmatrix} 0 & 1 \cr 1 & 0 \end{pmatrix} \, , \qquad
\sigma_{2} = \begin{pmatrix}0 & -i \cr i & \;\;\;0 \end{pmatrix} \, , \qquad
\sigma_{3} = \begin{pmatrix} 1 & \;\;\;0 \cr 0 & -1 \end{pmatrix} \, .
\end{equation}
The explicit forms of the operators in Equation~(\ref{eq:53}) are given in Table \ref{tabl1}. The $SL(2\, ,C)$ is homomorphic (the mapping is onto but not
one-to-one) to $SO(3\, ,1)$, namely it is the double cover of $SO(3\, ,1)$. 
\begin{table}[!t]
\caption{Generators and transformation matrices of $SL(2\,,C)$ and $SO(3\,,1)$. 
The transformation matrices of the Lorentz
group are applicable to the Minkowskian space of $(t, z, x, y)$.}\label{tabl1}
\centering
\noindent\hspace*{-\extralength}%
\begin{minipage}{\dimexpr\textwidth+\extralength\relax}
{\begin{tabular}{llll}
{}&{}&{}&{} \\
\hline\\[-2.4ex]
\hline\\
 Generators of & Transformation matrices & Generators of &  Transformation matrices\\
\hspace{3mm} $SL(2\,,C)$ & {} &\hspace{3mm}  $SO(3\,,1)$ & {}
  \\[0.8ex]
\hline \\[1.8ex]
  $J_{1} = \frac{1}{2}\begin{pmatrix}0 & 1 \cr 1 & 0 \end{pmatrix}$
&  
$R_{1}(\theta)=\begin{pmatrix}\cos(\theta/2) &  i\sin(\theta/2)
                  \cr i\sin(\theta/2) & \cos(\theta/2) \end{pmatrix}$  
 & 
 $J_{1}=\begin{pmatrix} 
0 & 0& 0 & 0 \\ 
0 & 0 & 0 & i \\ 
0 & 0 & 0& 0 \\
0 & -i & 0& 0 \end{pmatrix}$  
&   
$R_{x}(\theta) =\begin{pmatrix} 1 & 0 & 0 & 0 \cr 0 & \;\;\;\; \cos\theta & 0 & \sin\theta  \cr
   0 & 0 & 1 & 0 \cr  0 & -\sin\theta & 0 & \cos\theta \end{pmatrix}
  $ \\[7ex]
  $K_{1} = \frac{1}{2}\begin{pmatrix}0 & i \cr i & 0 \end{pmatrix}$
&  
$B_{1}(\lambda)=\begin{pmatrix}\cosh(\lambda/2) &  \sinh(\lambda/2)
       \cr \sinh(\lambda/2) & \cosh(\lambda/2) \end{pmatrix}$ 
&
 $K_{1}=\begin{pmatrix} 
0 & 0 & i & 0 \\
0 & 0& 0& 0 \\ 
i & 0& 0 & 0 \\ 
0 & 0 & 0 & 0 \end{pmatrix}$    
&  
$B_{x}(\lambda) =\begin{pmatrix}\cosh\lambda & 0 & \sinh\lambda & 0 \cr 0 & 1 & 0 & 0
 \cr
 \sinh\lambda & 0 & \cosh\lambda & 0 \cr 0 & 0 & 0 & 1 \end{pmatrix} $\\[7ex]
\hline\\ [-0.5ex]
 $J_{2} = \frac{1}{2}\begin{pmatrix}0 & -i \cr i & 0 \end{pmatrix}$
&  
$R_{2}(\theta)=\begin{pmatrix}\cos(\theta/2) &  -\sin(\theta/2)
       \cr \sin(\theta/2) & \;\;\;\; \cos(\theta/2) \end{pmatrix}$ 
&
$ J_{2}=\begin{pmatrix} 
 0 & 0 & \;\;\;0 & 0 \\
0& 0& -i & 0 \\
0 & i & \;\;\;0& 0 \\
0 & 0 & \;\;\;0 & 0 \end{pmatrix}$     
&  
$R_{y}(\theta)=\begin{pmatrix}1 & 0 & 0 & 0 \cr 0 & \cos\theta & -\sin\theta & 0 \cr
  0 & \sin\theta & \;\;\;\; \cos\theta & 0 \cr 0 & 0 & 0 & 1 \end{pmatrix} $ \\[7ex]
 $K_{2} = \frac{1}{2} \begin{pmatrix} 0 & 1 \cr -1 & 0 \end{pmatrix}$
&  
$B_{2}(\lambda)=\begin{pmatrix}\cosh(\lambda/2) &  -i\sinh(\lambda/2) \cr
               i\sinh(\lambda/2) & \;\;\;\; \cosh(\lambda/2) \end{pmatrix}$
& 
 $K_{2}=\begin{pmatrix} 
0 & 0 & 0 & i \\ 
0 & 0 & 0 & 0 \\
0 & 0& 0& 0 \\ 
i & 0 & 0& 0 \end{pmatrix} $            
&   
 $B_{y}(\lambda) =\begin{pmatrix}\cosh\lambda & 0 & 0 & \sinh\lambda  \cr 0 & 1 & 0 & 0 \cr
  0 & 0 & 1 & 0 \cr \sinh\lambda & 0 & 0 & \cosh\lambda \end{pmatrix} $ \\[5.5ex]
\hline\\ [-0.5ex]
 $J_{3} = \frac{1}{2}\begin{pmatrix}1 & 0 \cr 0  & -1 \end{pmatrix}$
&  

$R_{3}(\phi)=\begin{pmatrix}e^{i\theta/2} &  0
                  \cr 0 & e^{-i\theta/2} \end{pmatrix} $ 
 & 
 $J_{3}=\begin{pmatrix} 
0 & 0 & 0 & 0 \\
0 & 0 & 0 & 0\\ 
0 & 0 & 0   & -i \\
0 & 0 & i  & 0  \end{pmatrix}$                 
&
$R_{z}(\theta) =\begin{pmatrix}1 & 0 & 0 & 0 \cr 0 & 1 & 0 & 0 \cr
 0 & 0 & \cos\theta & -\sin\theta \cr
 0 & 0 & \sin\theta & \;\;\;\; \cos\theta \end{pmatrix} $ \\[7ex]
 $K_{3} = \frac{1}{2}\begin{pmatrix}i & 0 \cr 0 & -i \end{pmatrix}$
&  
$B_{3}(\lambda)=\begin{pmatrix}e^{\lambda/2} &  0 \cr 0 & e^{-\lambda/2} \end{pmatrix}$
&
$K_{3} = \begin{pmatrix} 
0 & i & 0& 0 \\ 
i &0 & 0& 0 \\ 
0& 0 & 0 & 0 \\
0& 0& 0 & 0 \end{pmatrix} $    
&  
$B_{z}(\lambda) =\begin{pmatrix}\cosh\lambda & \sinh\lambda & 0 & 0 \cr  \sinh\lambda & \cosh\lambda & 0 & 0 \cr
  0 & 0 & 1 & 0 \cr 0 & 0 & 0 & 1\end{pmatrix} $ \\ \\
\hline\\[-2.4ex]
\hline
\end{tabular}}
\end{minipage}
\end{table}
\FloatBarrier

\section{Generators of SO(3,2) and Sp(4)}
Dirac in 1963~\cite{dir63} showed that the generators of the Lorentz group $SO(3\,,2)$, which is isomorphic to the (3 + 2) de Sitter group, satisfy the following set of
commutation relations:
\begin{alignat}{2}\label{eq:9h310}
&[J_{i}, J_{j}] = i\varepsilon _{ijk} J_{k} \, ,\quad
&&[J_{i}, K_{j}] = i\varepsilon_{ijk} K_{k} \, , \nonumber\\[1ex]
&[J_{i}, Q_{j}] = i\varepsilon_{ijk} Q_{k} \, , \quad
&&[K_{i}, K_{j}] = [Q_{i}, Q_{j}] = -i\varepsilon _{ijk} J_{k} \, ,  \nonumber\\[1ex]
&[K_{i}, Q_{j}] = -i\delta_{ij} J_{0} , \quad &&[J_{i}, J_{0}] = 0
\, , \nonumber\\[1ex]
&[K_{i}, J_{0}] =  -iQ_{i} \, , \quad
&&[Q_{i}, J_{0}] = iK_{i} \, .
\end{alignat}
This $SO(3\, , 2)$ Lorentz group has ten generators and is
applicable to the three space and two time dimensions. A five-by-matrix representation of this Lie algebra can be found in~\cite{bkn21iop}, where it is also discussed how this group is important in dealing with space-time
symmetries. Below, Table \ref{tbl:6d51} gives the expressions for the generators of $Sp(4)$ in Fock space and the differential forms of the generators in the phase space of $(x_1,p_1,x_2,p_2).$

\begin{table}[htb]
\centering
\caption[Symmetries applicable to two mode squeezed states.]{Generators for $SO(3\,,2)$  and $Sp(4)$.  These generators are applicable to two mode squeezed states expressed in Fock space and in phase space, respectively.
They satisfy the same commutation relations.\\} \label{tbl:6d51}
\vspace{-2.5mm}
\begin{center}
\begin{tabular}{cccccc}
\hline
\hline\\[-2.0ex]
\hspace{1mm}& Generators &\hspace{1mm} & Fock space &\hspace{1mm}
  & Differential form in phase space  \\
\hline\\ [-1.0ex]
\hspace{1mm}&
$J_{1}$
&\hspace{1mm} &  $ \frac{1}{2}\left(\hat{a}^{\dagger}_{1}\hat{a}_{2} +
\hat{a}^{\dagger}_{2}\hat{a}_{1}\right) $ &\hspace{1mm} &
$ \frac{i}{2}\left\{\left(x_{1}\frac{\partial}{\partial p_{2}} - p_{2}\frac{\partial}{\partial x_{1}} \right) +
			\left(x_{2}\frac{\partial}{\partial p_{1}} - p_{1}\frac{\partial}{\partial x_{2}} \right)\right\} $
\\[1.5ex]
\hline\\ [-0.7ex] \hspace{1mm}& $J_{2} $
&\hspace{1mm} &
$\frac{1}{2i}\left(\hat{a}^{\dagger}_{1}\hat{a}_{2} - \hat{a}^{\dagger}_{2}\hat{a}_{1}\right)$
&\hspace{1mm} & $ \frac{i}{2}\left\{\left(x_{1}\frac{\partial}{\partial x_{2}} - x_{2}\frac{\partial}{\partial x_{1}} \right) +
			\left(p_{1}\frac{\partial}{\partial p_{2}} - p_{2}\frac{\partial}{\partial p_{1}} \right)\right\} $
\\[1.5ex]
\hline\\ [-0.7ex]
\hspace{1mm}&
$J_{3} $
&\hspace{1mm} &  
$ \frac{1}{2} \left(\hat{a}^{\dagger}_{1}\hat{a}_{1} - \hat{a}^{\dagger}_{2}\hat{a}_{2} \right) $
&\hspace{1mm} &
 $ -\frac{i}{2}\left\{\left(x_{1}\frac{\partial}{\partial p_{1}} - p_{1}\frac{\partial}{\partial x_{1}} \right)-
			\left(x_{2}\frac{\partial}{\partial p_{2}} - p_{2}\frac{\partial}{\partial x_{2}} \right)\right\} $
\\[1.5ex]
\hline\\ [-0.7ex]
\hspace{1mm}&
$J_{0}$
&\hspace{1mm} &  $\frac{1}{2}\left(\hat{a}^{\dagger}_{1}\hat{a}_{1} +
\hat{a}_{2}\hat{a}^{\dagger}_{2}\right)  $
&\hspace{1mm} & $ - \frac{i}{2}\left\{\left(x_{1}\frac{\partial}{\partial p_{1}} - p_{1}\frac{\partial}{\partial x_{1}} \right) +
			\left(x_{2}\frac{\partial}{\partial p_{2}} - p_{2}\frac{\partial}{\partial x_{2}} \right)\right\} $
\\[1.5ex]
\hline\\ [-0.7ex]
\hspace{1mm}&
$K_{1}$
&\hspace{1mm} &  $
 -\frac{1}{4}\left(\hat{a}^{\dagger}_{1}\hat{a}^{\dagger}_{1} + \hat{a}_{1}\hat{a}_{1} -
 \hat{a}^{\dagger}_{2}\hat{a}^{\dagger}_{2} - \hat{a}_{2}\hat{a}_{2}\right)$
&\hspace{1mm} & $ - \frac{i}{2}\left\{\left(x_{1}\frac{\partial}{\partial p_{1}} + p_{1}\frac{\partial}{\partial x_{1}} \right)
	        - \left(x_{2}\frac{\partial}{\partial p_{2}} + p_{2}\frac{\partial}{\partial x_{2}} \right)\right\} $
\\[1.5ex]
\hline\\ [-0.7ex]
\hspace{1mm}&
$K_{2}$
&\hspace{1mm} &  $ +\frac{i}{4}\left(\hat{a}^{\dagger}_{1}\hat{a}^{\dagger}_{1}
 - \hat{a}_{1}\hat{a}_{1} + \hat{a}^{\dagger}_{2}\hat{a}^{\dagger}_{2} - \hat{a}_{2}\hat{a}_{2}\right) $ 
&\hspace{1mm} & $ \frac{i}{2}\left\{\left(x_{1}\frac{\partial}{\partial x_{1}} + x_{2}\frac{\partial}{\partial x_{2}} \right)
			- \left(p_{1}\frac{\partial}{\partial p_{1}} + p_{2}\frac{\partial}{\partial p_{2}} \right)\right\} $
\\[1.5ex]
\hline\\ [-0.7ex]
\hspace{1mm}&
$K_{3}$
&\hspace{1mm} &
$ \frac{1}{2}\left(\hat{a}^{\dagger}_{1}\hat{a}^{\dagger}_{2} + \hat{a}_{1}\hat{a}_{2}\right)$
&\hspace{1mm} & $ - \frac{i}{2}\left\{\left(x_{1}\frac{\partial}{\partial p_{2}} + p_{2}\frac{\partial}{\partial x_{1}} \right)
			+ \left(x_{2}\frac{\partial}{\partial p_{1}} + p_{1}\frac{\partial}{\partial x_{2}} \right)\right\} $
\\[1.5ex]
\hline\\ [-0.7ex]
\hspace{1mm}&
$Q_{1}$
&\hspace{1mm} &  $
-\frac{i}{4}\left(\hat{a}^{\dagger}_{1}\hat{a}^{\dagger}_{1} - \hat{a}_{1}\hat{a}_{1} -
  \hat{a}^{\dagger}_{2}\hat{a}^{\dagger}_{2} + \hat{a}_{2}\hat{a}_{2} \right) $
&\hspace{1mm} & $ \frac{i}{2}\left\{\left(x_{1}\frac{\partial}{\partial x_{1}} - p_{1}\frac{\partial}{\partial p_{1}} \right)
		 	- \left(x_{2}\frac{\partial}{\partial x_{2}} - p_{2}\frac{\partial}{\partial p_{2}} \right)\right\}$
\\[1.5ex]
\hline\\ [-0.7ex]
\hspace{1mm}&
$Q_{2}$
&\hspace{1mm} &
 $ -\frac{1}{4}\left(\hat{a}^{\dagger}_{1}\hat{a}^{\dagger}_{1} + \hat{a}_{1}\hat{a}_{1} +
   \hat{a}^{\dagger}_{2}\hat{a}^{\dagger}_{2} + \hat{a}_{2}\hat{a}_{2} \right) $
&\hspace{1mm} & $  -\frac{i}{2}\left\{\left(x_{1}\frac{\partial}{\partial p_{1}} + p_{1}\frac{\partial}{\partial x_{1}} \right)
			+\left(x_{2}\frac{\partial}{\partial p_{2}} + p_{2}\frac{\partial}{\partial x_{2}} \right)\right\}   $
\\[1.5ex]
\hline\\ [-0.7ex]
  \hspace{1mm}&
$Q_{3}$ &\hspace{1mm} &  $
\frac{i}{2}\left(\hat{a}_{1}^{\dagger}\hat{a}_{2}^{\dagger} - \hat{a}_{1}\hat{a}_{2}\right) $
&\hspace{1mm} & $ - \frac{i}{2}\left\{\left(x_{2}\frac{\partial}{\partial x_{1}} + x_{1}\frac{\partial}{\partial x_{2}} \right)
			- \left(p_{2}\frac{\partial}{\partial p_{1}} + p_{1}\frac{\partial}{\partial p_{2}} \right)\right\} $ 
\\[2.0mm]
\hline
\hline\\[-0.4ex]
\end{tabular}
\end{center}
\end{table}

\FloatBarrier

\section*{Acknowledgment:} This work is dedicated to the memory of Young S. Kim of the University of Maryland, College Park, 
in recognition of his many years of collaboration filled with insights and ideas.


\bibliography{References}

\end{document}